\documentclass[12pt,preprint]{aastex}
\usepackage{graphicx}

\begin{document}

\title{New cooling sequences for old white dwarfs}

\author{I. Renedo$^{1,2}$,
	L. G. Althaus$^{1,3}$,
        M. M. Miller Bertolami$^{3,4}$,
        A. D. Romero$^{3,4}$,
        A. H. C\'orsico$^{3,4}$,
        R. D. Rohrmann$^{5}$,
        and \\
        E. Garc\'{\i}a--Berro$^{1,2}$}

\affil{$^1$Departament de F\'\i sica Aplicada, 
           Universitat Polit\`ecnica de Catalunya,
           c/Esteve Terrades 5, 
           08860 Castelldefels,
           Spain\\
       $^2$Institute for Space Studies of Catalonia,
           c/Gran Capit\`a 2--4, 
           Edif. Nexus 104, 
           08034 Barcelona,
           Spain\\
       $^3$Facultad de Ciencias Astron\'omicas y Geof\'{\i}sicas,
           Universidad Nacional de La Plata, 
           Paseo del Bosque s/n, 
           (1900) La Plata, 
           Argentina\\
       $^4$Instituto de Astrof\'{\i}sica de La Plata,
           IALP (CCT La Plata), 
           CONICET-UNLP\\
       $^5$Observatorio Astron\'omico,
           Universidad Nacional de C\'ordoba, 
           Laprida 854, 
           (5000) C\'ordoba, 
           Argentina}

\email{garcia@fa.upc.edu}

\begin{abstract} 
We present full evolutionary calculations appropriate for the study of
hydrogen-rich DA white  dwarfs.  This is done by  evolving white dwarf
progenitors from the zero age main sequence, through the core hydrogen
burning  phase, the  helium burning  phase and  the  thermally pulsing
asymptotic  giant branch  phase to  the white  dwarf  stage.  Complete
evolutionary sequences are computed for a wide range of stellar masses
and for two different metallicities: $Z=0.01$, which is representative
of the solar neighborhood, and $Z=0.001$, which is appropriate for the
study of old stellar systems, like globular clusters. During the white
dwarf cooling  stage we compute  self-consistently the phase  in which
nuclear reactions are still  important, the diffusive evolution of the
elements in the  outer layers and, finally, we  also take into account
all  the relevant energy  sources in  the deep  interior of  the white
dwarf,  like   the  release  of   latent  heat  and  the   release  of
gravitational  energy  due  to  carbon-oxygen  phase  separation  upon
crystallization.   We also  provide  colors and  magnitudes for  these
sequences, based on  a new set of improved  non-gray white dwarf model
atmospheres, which  include the  most up-to-date physical  inputs like
the Ly$\alpha$ quasi-molecular opacity.  The calculations are extended
down to an effective temperature  of 2,500 K. Our calculations provide
a homogeneous set of  evolutionary cooling tracks appropriate for mass
and  age determinations of  old DA  white dwarfs  and for  white dwarf
cosmochronology of the different Galactic populations.
\end{abstract}         



\section{Introduction}
\label{intro}

White dwarfs are the most common stellar evolutionary end-point.  As a
matter of  fact, more  than 97\%  of stars are  expected to  end their
lives as  white dwarfs. Their evolution  can be described  as a simple
cooling process which lasts for  very long periods of time.  Moreover,
since  these fossil  stars are  abundant and  long-lived  objects they
convey  important information  about  the properties  of all  galactic
populations (Hansen \& Liebert  2003). In particular, white dwarfs can
be considered  as reliable cosmic  clocks to infer  the age of  a wide
variety of  stellar populations,  such as the  Galactic disk  and halo
(Winget  et al.   1987; Garc\'\i  a--Berro et  al.  1988a,b;  Isern et
al. 1998; Garc\'\i a--Berro et al. 1999; Torres et al.  2002), and the
system of globular  and open clusters (Kalirai et  al. 2001; Hansen et
al. 2002, 2007).  They also provide helpful information about the past
history of the  star formation rate of our  Galaxy (D\'\i az--Pinto et
al.  1994).   Additionally, they can  be used to place  constraints on
exotic elementary particles (Isern et al. 1992; C\'orsico et al. 2001;
Isern et al. 2008) or on alternative theories of gravitation (Garc\'\i
a--Berro et al.  1995; Benvenuto  et al. 2004).  Recent reviews on the
properties and evolution of white dwarfs and of their applications are
those  of Winget  \& Kepler  (2008) and  Fontaine \&  Brassard (2008).
However, to use them for  this wide range of applications reliable and
accurate evolutionary sequences must be computed.

Previous  recent  evolutionary  calculations  of  hydrogen-rich  white
dwarfs  are those of  Fontaine et  al. (2001),  Salaris et  al. (1997;
2000),  Hansen (1998;  1999),  Benvenuto \&  Althaus  (1999) and  Wood
(1992; 1995).  All  these works have studied different  aspects of the
evolution of white dwarfs with hydrogen-rich envelopes.  For instance,
the  most commonly used  models, those  of Wood  (1995), cover  a wide
range of stellar masses and  envelope masses and, until recently, were
considered  to be a  standard reference  in the  field of  white dwarf
cosmochronology.   However,  we   emphasize  that  these  models  were
computed using gray atmospheres,  a severe drawback (especially at low
luminosities) that more recent calculations have overcome. Among these
models  we mention  the works  of Hansen  (1998; 1999).   This  set of
cooling models  pioneered the usage  of detailed model  atmospheres as
surface  boundary conditions  in  cooling calculations  for old  white
dwarfs. This  is an important issue  since it affects  the location of
the base  of the convective  envelope. These calculations  also showed
that collision-induced  absorption processes affect the  colors of old
white dwarfs. Salaris  et al. (1997; 2000) focused  on the question of
the interior  abundances of  carbon and oxygen,  which is  of critical
importance  to  derive reliable  ages  and  on  the effects  of  phase
separation of  carbon and oxygen upon  crystallization, while Fontaine
et al.  (2001) were the first to discover the importance of convective
coupling between  the atmosphere and  the degenerate core.   This last
issue  also  bears  importance   for  the  determination  of  accurate
ages. Nevertheless, all these  works suffer from the same shortcoming.
All of them  evolved intial white dwarf configurations  which were not
obtained   self-consistently    from   models   evolving    from   the
main-sequence.  As  a consequence, the chemical  stratification of the
inner, degenerate core was simplistic  in most of the cases, except in
the case  of the  cooling sequences of  Salaris et al.   (1997; 2000).
Also, the envelope mass and the outer layer chemical distribution were
idealized ones  in all the cases.  Additionally,  the oldest sequences
used physical inputs which are nowadays outdated and, finally, most of
them disregarded the energy  release of carbon-oxygen phase separation
(Garc\'\i a--Berro et al. 1988a; 1988b).

The aim of this paper is to compute a set of new cooling sequences for
hydrogen-rich white dwarfs, incorporating the most up-to-date physical
inputs.  We emphasize that our evolutionary sequences are derived from
a  full and  self-consistent  treatment of  the complete  evolutionary
history  of progenitor stars  with different  masses evolved  with two
different metallicities  ($Z=0.01$ and $Z=0.001$)  appropriate for the
study  of the solar  neighborhood and  of metal-poor  stellar systems,
like globular  clusters or the galactic halo.   Thus, our calculations
constitute a  comprehensive set of evolutionary tracks  which allow to
study the evolution of hydrogen-rich white dwarfs in a wide variety of
stellar  systems.  Moreover,  since  our  calculations  encompass  the
pre-white  dwarf  evolutionary phases,  the  white dwarf  evolutionary
calculations  presented  here  are  not  affected  by  inconsistencies
arising from  artificial procedures  to generate starting  white dwarf
configurations.   In particular, the  calculation of  the evolutionary
history of progenitor  stars provides us with the amount  of H left in
the white dwarf, and with  the chemical profiles expected not only for
the carbon-oxygen core, but  also for the partially degenerate regions
above the core, of relevance for the white dwarf cooling phase.

The paper  is organized as follows.  In  Sect.  \ref{Computational} we
describe the main physical inputs of our models.  We also describe the
main  characteristics of  the  initial models,  the model  atmospheres
employed  in   this  work  and   some  details  of   the  evolutionary
computations.  In  Sect.  3  we present the  results of  the pre-white
dwarf evolutionary sequences, paying  special attention to the derived
initial-to-final-mass relationship,  and also the  white dwarf cooling
tracks. In this section we also discuss in detail the effects of phase
separation upon crystallization.  Finally, in Sect. 4 we summarize our
results, we discuss their significance and we draw our conclusions.


\section{Computational details}
\label{Computational}

\subsection{Input physics}
\label{Input physics}

The evolutionary calculations presented in this work were done with an
updated version of the {\tt  LPCODE} stellar evolutionary code --- see
Althaus et al.  (2005a) and references therein.  We  used this code to
compute  both  the  evolution  of  white  dwarfs  and  that  of  their
progenitor stars.  We emphasize that in recent years, the {\tt LPCODE}
stellar evolutionary code has been employed to study different aspects
of  the  evolution  of  low-mass  stars, such  as  the  formation  and
evolution of H-deficient white  dwarfs, PG 1159 and extreme horizontal
branch stars (Althaus et al.  2005a; Miller Bertolami \& Althaus 2006;
Miller  Bertolami et  al.   2008;  Althaus et  al.   2009a) and,  more
recently, it has also been used to study the formation of hot DQ white
dwarfs (Althaus et al.  2009b),  the evolution of He-core white dwarfs
with  high metallicity progenitors  (Althaus et  al.  2009c),  and the
evolution of hydrogen-deficient white  dwarfs (Althaus et al.  2009d).
Moreover,  this code  has  also been  used  to study  the white  dwarf
initial-final mass relationship (Salaris  et al. 2009).  A recent test
of the  code and a thorough  comparison of the  results obtained using
{\tt LPCODE}  with those obtained  using other evolutionary  codes has
recently been made in Salaris  et al. (2009).  Details of {\tt LPCODE}
can be found in these works  and in Althaus et al. (2009d).  For these
reasons, in what  follows, we only comment on  the main physical input
physics,  namely,  those  that   are  relevant  for  the  evolutionary
calculations presented in this work.

To begin with, we recall  that the radiative opacities employed in our
calculations were those of the OPAL project (Iglesias \& Rogers 1996),
including  carbon- and oxygen-rich  compositions, supplemented  at low
temperatures   with  the  Alexander   \&  Ferguson   (1994)  molecular
opacities.   For  the present  calculations,  we  have not  considered
carbon-enriched molecular opacities  (Marigo 2002), which are expected
to  reduce  effective  temperatures  at  the AGB  (Weiss  \&  Ferguson
2009). We adopted  the conductive opacities of Cassisi  et al. (2007),
which cover  the entire regime where electron  conduction is relevant.
Neutrino emission rates for  pair, photo, and bremsstrahlung processes
were  those of  Itoh et  al.  (1996),  while for  plasma  processes we
included the treatment presented in Haft et al.  (1994).

{\tt LPCODE}  considers a simultaneous  treatment of non-instantaneous
mixing and  burning of  elements. Specifically, abundance  changes are
described by the set of equations

\begin{equation} \label{ec1}
\left( \frac{d \vec{X}}{dt} \right) = 
\left( \frac{\partial \vec{X}}{\partial t} \right)_{\rm nuc} +
\frac{\partial}{\partial M_r} \left[ (4\pi r^2 \rho)^2 D 
\frac{\partial \vec{X}}{\partial M_r}\right], 
\label{mezcla}
\end{equation} 

\noindent where $\vec{X}$ is a vector containing the abundances of all
the  elements  --- see  Althaus  et al.   (2003)  for  details on  the
numerical procedure.   In this equation, the first  term describes the
nuclear evolution, and it is  fully coupled to the current composition
change due to mixing processes,  as represented by the second term. In
particular, the  efficiency of convective  mixing is described  by the
diffusion coefficient $D$,  which in this work is  given by the mixing
length  theory.   The  nuclear  network accounts  explicitly  for  the
following 16 elements: $^{1}$H, $^{2}$H, $^{3}$He, $^{4}$He, $^{7}$Li,
$^{7}$Be, $^{12}$C, $^{13}$C,  $^{14}$N, $^{15}$N, $^{16}$O, $^{17}$O,
$^{18}$O,  $^{19}$F,   $^{20}$Ne  and  $^{22}$Ne,   together  with  34
thermonuclear reaction  rates of the  pp chains, CNO  bi-cycle, helium
burning, and carbon ignition that  are identical to those described in
Althaus et al.  (2005a), with the exception of the reaction $^{12}$C$\
+\ p \rightarrow  \ ^{13}$N + $\gamma \rightarrow \ ^{13}$C  + $ e^+ +
\nu_{\rm e}$ and $^{13}$C(p, $\gamma)^{14}$N, for which we adopted the
rate of Angulo et al.  (1999).  In particular, it is worth noting that
the  $^{12}$C($\alpha,\gamma)^{16}$O  reaction rate  ---  which is  of
special relevance for the carbon-oxygen stratification of the emerging
white  dwarf core  --- adopted  here was  also that  of Angulo  et al.
(1999). In  passing, we mention  that a detailed inner  composition is
required  for a  proper  computation  of the  energy  released by  the
redistribution  of  chemical elements  during  crystallization of  the
white dwarf core (see below). Specifically, the energy released during
this process,  and the resulting impact  on the cooling  ages of faint
white  dwarfs,  increases  for  larger  carbon  abundances  (Isern  et
al. 1997; 2000).  The standard mixing length theory for convection ---
with  the free  parameter  $\alpha=1.61$ ---  was  adopted. With  this
value, the  present luminosity and  effective temperature of  the Sun,
$\log T_{\rm  eff}= 3.7641$ and  $L_{\sun}= 3.842 \times  10^{33}$ erg
s$^{-1}$, at an  age of 4570 Myr, are reproduced  by {\tt LPCODE} when
$Z=0.0164$ and $X=0.714$  are adopted --- in agreement  with the $Z/X$
value of Grevesse \& Sauval (1998).

During  the evolutionary  stages prior  to the  thermally  pulsing AGB
(TP-AGB)  phase, we  allowed the  occurrence of  extra-mixing episodes
beyond each  convective boundary following the  prescription of Herwig
et al.  (1997). As well known, the occurrence of extra-mixing episodes
is of  relevance for  the final chemical  stratification of  the white
dwarfs,  particularly during  the late  stage of  core  helium burning
phase ---  see Prada Moroni \&  Straniero (2002) and  Straniero et al.
(2003).  We treated extra-mixing as a time-dependent diffusion process
--- by assuming that mixing velocities decay exponentially beyond each
convective boundary --- with  a diffusion coefficient given by $D_{\rm
EM}=  D_{\rm O}\  \exp(-2z/f H_{\rm  p})$,  where $H_{\rm  P}$ is  the
pressure scale height  at the convective boundary, $D_{\rm  O}$ is the
diffusion  coefficient of  unstable  regions close  to the  convective
boundary,  and $z$  is the  geometric distance  from the  edge  of the
convective boundary (Herwig  et al.  1997).  For all  our sequences we
adopted $f=0.016$, a  value inferred from the width  of the upper main
sequence.   In the case  of the  sequence of  $1.25 \,  M_{\sun}$ with
$Z=0.001$, the  size of  the convective core  on the main  sequence is
small.  In this case we use $f=0.008$ during the core hydrogen burning
phase.   Abundance  changes  resulting  from extra-mixing  were  fully
coupled to nuclear evolution, following Eq. (\ref{mezcla}).

In the  present study,  extra-mixing episodes were  disregarded during
the TP-AGB  phase. In particular, a  strong reduction (a  value of $f$
much smaller than  0.016) of extra-mixing episodes at  the base of the
pulse-driven convection  zone seems to be  supported by 
simulations  of  the $s-$process  abundance  patterns  (Lugaro et  al.
2003) and,   more  recently,  by   observational  inferences   of  the
initial-final mass relation (Salaris et al.  2009). As a result, it is
expected  that the  mass of  the hydrogen-free  core of  our sequences
gradually  grows as evolution  proceeds through  the TP-AGB.   This is
because a strong reduction or  suppression of extra-mixing at the base
of the  pulse-driven convection zone strongly  inhibits the occurrence
of  third dredge-up,  thus favoring  the growth  of  the hydrogen-free
core.   The  implications  of   this  treatment  for  the  theoretical
initial-final mass relation will be  discussed later in this paper ---
see also  Salaris et  al.  (2009) and  Weiss \& Ferguson  (2009).  The
breathing pulse  instability occurring towards the end  of core helium
burning  was  suppressed  ---  see  Straniero et  al.   (2003)  for  a
discussion of this point.

Mass  loss was  considered during  core helium  burning and  red giant
branch phases  following Schr\"oder \&  Cuntz (2005).  During  the AGB
and TP-AGB  phases, we considered  the prescription of  Vassiliadis \&
Wood (1993).  In the case of a strong reduction of third dredge-up, as
occurred in our sequences, mass loss plays a major role in determining
the final  mass of  the hydrogen-free  core at the  end of  the TP-AGB
evolution, and  thus the  initial-final mass relation.  In particular,
mass loss erodes the hydrogen-rich envelope of the star and limits the
additional growth  of the core mass  during the TP-AGB.  This is quite
different  from the  situation  in which  appreciable third  dredge-up
takes place, in which case the  final core mass at the TP-AGB phase is
not very different from the mass  at the first thermal pulse (Weiss \&
Ferguson 2009), and the role of mass loss becomes less relevant.

We want  to mention that {\tt  LPCODE} is able to  follow the complete
evolution of  the star model from  the main sequence  through the core
helium flash and,  finally, for many thermal pulses  on the AGB almost
without hand intervention, except for  the latest stages of the TP-AGB
phase  of the  more massive  sequences, where  numerical instabilities
arise.  These instabilities,  also  found by  other  authors, are  the
result of the large radiation pressure in the convective envelope.  To
circumvent them,  we found computationally  convenient to artificially
modify the opacity  profile in those regions close to  the base of the
convective  envelope.   This  procedure  bears no  relevance  for  the 
evolution during the white dwarf regime.

For the  white dwarf  regime, the physical  inputs considered  in {\tt
LPCODE} were  completely revised and updated from  those considered in
our  previous   studies,  particularly  regarding   the  treatment  of
crystallization.    With   regard  to   the   microphysics,  for   the
high-density regime, we used the equation of state of Segretain et al.
(1994), which  accounts for all  the important contributions  for both
the  liquid  and solid  phases  --- see  Althaus  et  al.  (2007)  and
references therein.   For the low-density  regime, we used  an updated
version  of the  equation  of  state of  Magni  \& Mazzitelli  (1979).
Radiative  and  conductive  opacities   are  those  described  at  the
beginning of  this section.   In particular, conductive  opacities are
considered for densities  larger than that at which  the main chemical
constituents are completely  pressure-ionized.  During the white dwarf
regime, the metal mass fraction $Z$  in the envelope is not assumed to
be  fixed,  rather,  it  is  specified consistently  acording  to  the
prediction of  element diffusion.  To account for  this, we considered
radiative opacities tables from OPAL for arbitrary metallicities.  For
effective  temperatures less than  10,000K we  include the  effects of
molecular  opacity  by assuming  pure  hydrogen  composition from  the
computations  of  Marigo  \&   Aringer  (2009).   This  assumption  is
justified because  element diffusion leads to  pure hydrogen envelopes
in  cool white  dwarfs.  For  the white  dwarf regime,  convection was
treated in the formalism of the  mixing length theory, as given by the
ML2 parameterization (Tassoul et al. 1990).

As well  known, there are  several physical processes that  change the
chemical abundance distribution of white dwarfs along their evolution.
In  particular,  element  diffusion  strongly  modifies  the  chemical
composition profile  throughout their outer  layers. In this  work, we
computed  the white  dwarf  evolution  in a  consistent  way with  the
changes of chemical abundance distribution caused by element diffusion
along the entire cooling phase.   As a result, our sequences developed
pure hydrogen envelopes, the thickness of which gradually increases as
evolution proceeds.  We  considered gravitational settling and thermal
and chemical diffusion --- but not radiative levitation, which is only
relevant at  high effective  temperatures for determining  the surface
composition --- of $^1$H, $^3$He, $^4$He, $^{12}$C, $^{13}$C, $^{14}$N
and $^{16}$O, see Althaus et  al.  (2003) for details.  In particular,
our   treatment  of   time-dependent   diffusion  is   based  on   the
multicomponent  gas treatment  presented in  Burgers (1969).   In {\tt
LPCODE}, diffusion becomes operative once the wind limit is reached at
high effective temperatures (Unglaub \& Bues 2000).  We assume this to
occur when the surface gravity of our models $g > g_{\rm diff}$, where
$g_{\rm diff}= 7$ if $T_{\rm eff}  > 90,000K$ and $g_{\rm diff}= 6.4 +
T_{\rm  eff} / 150,000  K$ if  $T_{\rm eff}  < 90,000K$.   For smaller
gravities,  wind mass-loss  is high  enough that  prevents appreciable
element  diffusion  from   occurring.   This  prescription  represents
reasonably well the detailed simulations of Unglaub \& Bues (2000) for
the occurrence  of wind limits  in hydrogen-rich white  dwarfs.  Other
physical process  responsible for changes  in the chemical  profile of
white dwarfs  that we  took into account  is related  to carbon-oxygen
phase separation  during crystallization. In this  case, the resulting
release of  gravitational energy considerably impacts  the white dwarf
cooling  times.   Abundance changes  resulting  from residual  nuclear
burning --- mostly during the  hot stages of white dwarf evolution ---
and  convective   mixing,  were  also   taken  into  account   in  our
simulations.  In  particular, the release of energy  by proton burning
was  considered down to  $\log(L/L_{\sun}) \approx  -4$.  The  role of
residual  hydrogen burning  in evolving  white dwarfs  is by  no means
negligible, particularly  in the case of those  white dwarfs resulting
from low-metallicity progenitors.  Finally, we considered the chemical
rehomogenization  of  the   inner  carbon-oxygen  profile  induced  by
Rayleigh-Taylor instabilities following Salaris et al.  (1997).  These
instabilities  arise because the  positive molecular  weight gradients
that remain above the flat  chemical profile left by convection during
helium core burning.
 
An important  aspect of the present  work was the  inclusion of energy
sources resulting  from the crystallization  of the white  dwarf core.
This  comprises  the  release  of  latent  heat  and  the  release  of
gravitational  energy  associated with  changes  in the  carbon-oxygen
profile induced  by crystallization.  In this study,  the inclusion of
these  two  energy  sources  was done  self-consistently  and  locally
coupled to the  full set of equations of  stellar evolution.  That is,
we  computed the  structure and  evolution  of white  dwarfs with  the
changing  composition   profile  and  with   the  luminosity  equation
appropriately modified  to account for both the  local contribution of
energy released from the core chemical redistribution and latent heat.
This constitutes an improvement over previous attempts (Salaris et al.
2000) to  include  the release  of  energy  from  phase separation  in
stellar evolutionary  codes. Details about the  numerical procedure to
compute the energy sources from crystallization will be presented in a
forthcoming work.  Briefly, at  each evolutionary timestep we computed
the crystallization temperature and the change of chemical composition
resulting from  phase separation using the  spindle-type phase diagram
of  Segretain \&  Chabrier (1993).   This phase  diagram  provides the
crystallization temperature (which depends on the chemical composition
of the liquid phase) as  a function of the crystallization temperature
of a  one component  plasma.  In our  calculations, the  one component
plasma  crystallization temperature  is computed  by  imposing $\Gamma
=180$,  where  $\Gamma  \equiv  \langle Z^{5/3}\rangle  e^2/a_{\rm  e}
k_{\rm  B}T$ is  the ion  coupling constant,  and $a_{\rm  e}$  is the
interelectronic distance. After computing the chemical compositions of
both the solid and liquid  phases we evaluated the net energy released
in the process as in Isern et  al.  (1997), and added to it the latent
heat contribution, of the order of $0.77 k_{\rm B}T$ per ion, which is
usually  smaller. Both  energy contributions  were distributed  over a
small mass  range around the  crystallization front.  We  mention that
the magnitude of both energy  sources was calculated at each iteration
during the convergence of the model.

\subsection{Model atmospheres}

A proper treatment of the  evolutionary behavior of cool white dwarfs
requires the use of outer  boundary conditions as provided by detailed
non-gray  model  atmospheres.   To   this  end,  we  considered  model
atmospheres that incorporate non-ideal  effects in the gas equation of
state  and  chemical  equilibrium, collision-induced  absorption  from
molecules, and the  Ly$\alpha$ quasi-molecular opacity.  Specifically,
for  $T_{\rm  eff} <  10,000K$,  we  derived  starting values  of  the
pressure, temperature, radial thickness  and outer mass fraction at an
optical  depth $\tau=25$  from a  grid of  non-gray  model atmospheres
which covers a surface gravity range between $\log g= 6.5$ and 9.  For
larger  values of  $\tau$,  the  use of  Rosseland  mean opacities  is
justified, and the diffusion  approximation for the radiative transfer
can be assumed.  The use of non-gray model atmospheres to derive outer
boundary conditions gives rise to shallower outer convection zones, as
compared with the standard  gray treatment of the atmosphere (Bergeron
et  al.  1997).   At advanced  stages  of white  dwarf evolution,  the
central   temperature  becomes  strongly   tied  to   the  temperature
stratification  of  the  outer  layers.  Thus,  using  non-gray  model
atmospheres is  highly desired for  an accurate assessment  of cooling
times of cool white dwarfs (Prada Moroni \& Straniero 2007).

In the  present work, model atmospheres were  specifically computed on
the  basis of improved  LTE model  atmospheres. Colors  and magnitudes
were evaluated for effective  temperatures lower than 60,000K, because
NLTE  effects become  important above  this  temperature. Calculations
were done for a pure hydrogen  composition and for the HST ACS filters
(Vega-mag system) and $UBVRI$  photometry.  The numerical code used is
a new  and updated  version of  the one described  in Rohrmann  et al.
(2002).     Models   were    computed    assuming   hydrostatic    and
radiative-convective equilibrium. Convective  transport present in the
cooler atmospheres  was treated  within the usual  mixing-length (ML2)
approximation,  adopting  the  same  value  of $\alpha$  used  in  the
evolutionary  calculations.  The  microphysics included  in  the model
atmospheres comprises  non-ideal effects in the gas  equation of state
and chemical equilibrium based on the occupation probability formalism
as described in Rohrmann et  al.  (2002).  The code includes H, H$_2$,
H$^+$,  H$^-$,  H$_2^+$,   H$_3^+$,  He,  He$^-$,  He$^+$,  He$^{2+}$,
He$_2^+$, HeH$^+$, and e$^-$.   The level occupation probabilities are
self-consistently  incorporated in  the  calculation of  the line  and
continuum opacities. Collision-induced absorptions due to H$_2$-H$_2$,
H$_2$-He,  and H-He  pairs are  also taken  into account  (Rohrmann et
al. 2002).

For the purpose of the  present work, the model atmospheres explicitly
included  the  Ly$\alpha$  quasi-molecular  opacity according  to  the
approximation  used  by Kowalski  \&  Saumon (2006).   Quasi-molecular
absorption   results   from  perturbations   of   hydrogen  atoms   by
interactions  with  other particles,  mainly  H  and  H$_2$. Here,  we
considered   extreme  pressure-broadening   of  the   line  transition
H($n=1$)$\rightarrow$H($n=2$) due to  H-H and H-H$_2$ collisions, with
the  red wing  extending  far  into the  optical  region.  A  detailed
description of  these collisional line-broadening  evaluations will be
presented in a forthcoming paper.   On the basis of the approximations
outlined  in Kowalski  \& Saumon  (2006),  we evaluated  the red  wing
absorption   within  the   quasi-static  approach   using  theoretical
molecular potentials to describe  the interaction between the radiator
and  the   perturber.   We  also  considered  the   variation  in  the
electric-dipole  transition moment  with  the interparticle  distance.
The H$_3$ energy-potential surfaces contributing to collisions H-H$_2$
were  taken  from  Kulander \& Guest (1979) and Roach \& Kuntz (1986), and the  dipole transition  moments were  calculated from
Petsalakis  et  al. (1988).   Broadening  of  Ly$\alpha$  line by  H-H
collisions  plays a  minor role  compared to  H-H$_2$  encounters. The
potential interactions for H-H configurations were taken from Kolos \&
Wolniewicz (1965) and the  transition probability was assumed constant
in  this case.   The  main effect  of  the Ly$\alpha$  quasi-molecular
opacity is  a reduction  of the predicted  flux at  wavelength smaller
than $4000$  \AA\ for  white dwarfs cooler  than $T_{\rm  eff} \approx
6,000$~K.

\subsection{Initial models}

The initial models for our white dwarf sequences correspond to stellar
configurations  derived  from the  full  evolutionary calculations  of
their progenitor stars. The initial  He content of our starting models
at the main sequence were provided by the relation $Y= 0.23 + 2.41 Z$,
as given  by present determinations  of the chemical evolution  of the
Galaxy (Flynn  2004; Casagrande et al.  2007).   Two metallicities for
the progenitor  stars were  considered: $Z=0.01$ and  $0.001$.  Hence,
the   initial  compositions  of   our  sequences   are,  respectively,
$(Y,Z)=(0.254,0.01)$  and  $(Y,Z)=(0.232,0.001)$.   All the  sequences
were evolved from the ZAMS through the thermally-pulsing and mass-loss
phases  on the AGB  and, finally,  to the  white dwarf  cooling phase.
Specifically, we computed 10  full evolutionary sequences for $Z=0.01$
and  six  for $Z=0.001$.  In  Table  \ref{table_MiMf2Z},  we list  the
initial  masses of  the progenitor  stars at  the ZAMS,  together with
other evolutionary quantities which will be discussed below. 


\section{Evolutionary results}
\label{results}

\subsection{From the ZAMS to the white dwarf stage}

\begin{figure}
\begin{center}
\includegraphics[clip,width=0.9\columnwidth,angle=-90]{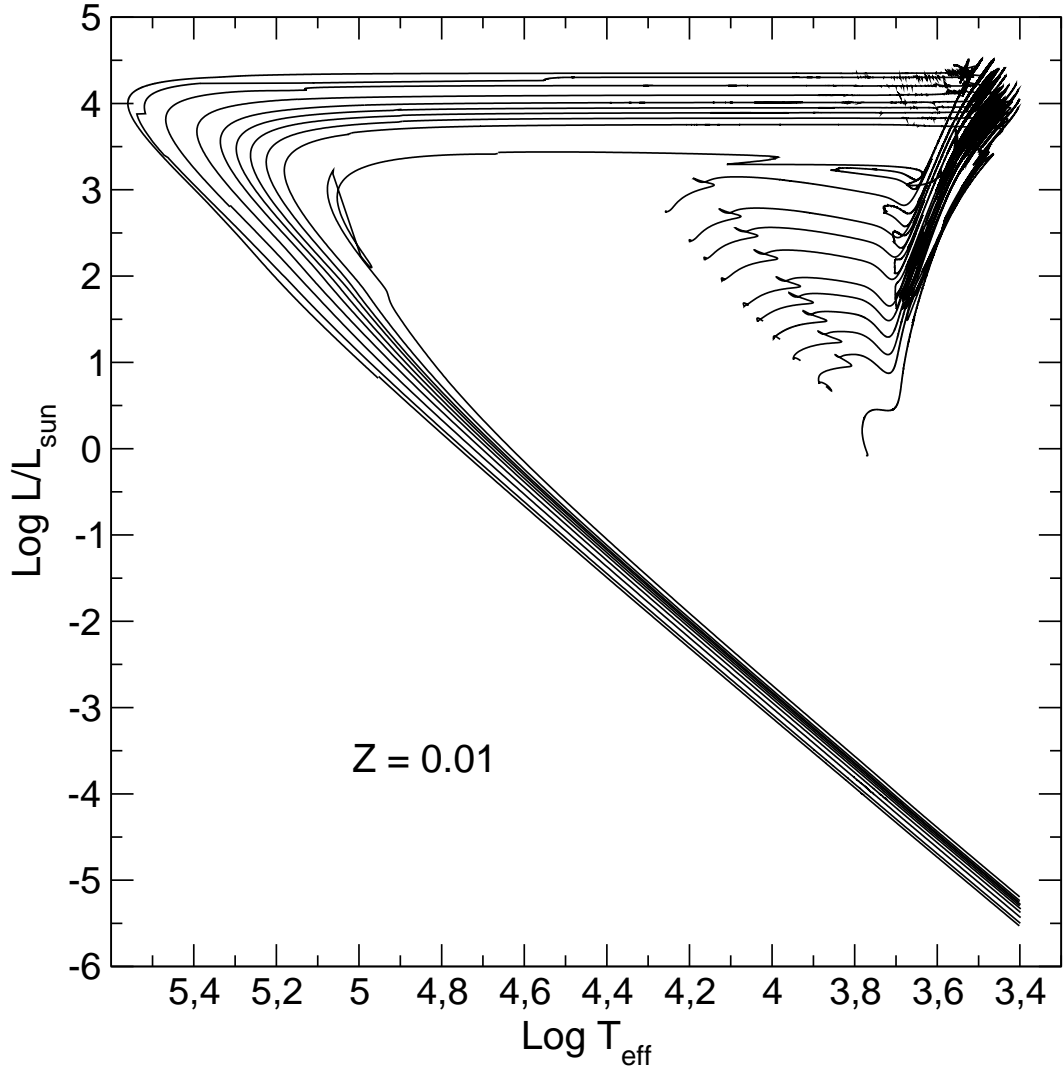}
\end{center}
\caption{Hertzsprung-Russell diagram of our evolutionary sequences for
         $Z=0.01$.   From  bottom  to  top: evolution  of  the  $1.0\,
         M_{\sun}$,   $1.5\,  M_{\sun}$,  $1.75\,   M_{\sun}$,  $2.0\,
         M_{\sun}$,   $2.25\,  M_{\sun}$,  $2.5\,   M_{\sun}$,  $3.0\,
         M_{\sun}$,  $3.5\,  M_{\sun}$,  $4.0\, M_{\sun}$  and  $5.0\,
         M_{\sun}$  model stars.  Evolutionary  tracks are  shown from
         the ZAMS  to advanced stages  of white dwarf  evolution. Note
         that  the  less   massive  sequence  experiences  a  hydrogen
         subflash before entering  its final cooling track.}
\label{HR_Zsolar}
\end{figure}

The evolution in the Hertzsprung-Russell diagram of our sequences from
the ZAMS to advances stages of  white dwarf evolution is shown in Fig.
\ref{HR_Zsolar}, for the case in which $Z=0.01$ is adopted.  Note that
the  less  massive sequence  experiences  a  hydrogen subflash  before
entering its final cooling track.   The initial masses at the ZAMS and
the final white dwarf masses of  these sequences can be found in Table
\ref{table_MiMf2Z},  for both  metallicities.  In  this table  we also
list  the   main-sequence  lifetimes,   which,  as  well   known,  for
solar-metallicity sequences are larger  than those of their metal-poor
counterparts. We mention that  our models have main-sequence lifetimes
longer  than those  recently published  by Weiss  \&  Ferguson (2009).
Differences are less  than 8 \% however, except  for the main-sequence
lifetime of  the solar  sequence, which  is 25 \%  larger. One  of the
reason for  such a discrepancy is  due to our  simplified treatment of
the equation  of state during  the evolutionary stages prior  to white
dwarf formation.   Also listed in this  table are the  total masses of
the residual hydrogen  and helium content left in  the white dwarfs at
the evolutionary stage corresponding to the point of maximum effective
temperature  in  the   Hertzsprung-Russell  diagram.   Note  that  the
residual  hydrogen content  decreases  with the  white  dwarf mass,  a
well-known result. For the  case of solar metallicity progenitors, the
hydrogen mass  differs by a  factor of 20  for the stellar  mass range
considered. This  general trend is also  observed for the  mass of the
helium  content, where  the mass  of the  residual helium  ranges from
0.025 to $0.0022 \, M_{\sun}$.   The hydrogen and helium masses listed
in Table  \ref{table_MiMf2Z} should be considered as  upper limits for
the  maximum  hydrogen  and  helium  content left  in  a  white  dwarf
resulting from the evolution  of single star progenitors. However, the
occurrence of a late thermal pulse after departure from the TP-AGB may
reduce the hydrogen mass considerably, see Althaus et al. (2005b).

Note  also  that for  a  given  white dwarf  mass  there  is a  marked
dependence of  the final hydrogen  mass on the initial  metallicity of
the progenitor star: higher hydrogen masses are expected in metal-poor
progenitors, see  Iben \& MacDonald  (1985, 1986).  For  instance, for
the $0.593  \, M_{\sun}$ white dwarf  model we find  that the hydrogen
mass  is  $\log M_{\rm  H}\simeq  -3.950$  for  the solar  metallicity
progenitor,  while this  mass turns  out to  be $\log  M_{\rm H}\simeq
-3.777$, for the  metal-poor progenitor, i.e. a 50\%  higher.  This is
an important issue since one  of the factors affecting the white dwarf
cooling  rate  is,  precisely,  the  thickness  of  the  hydrogen-rich
envelope. By contrast, in the  case of the residual helium content, no
appreciable dependence on the metallicity exists.  Also shown in Table
\ref{table_MiMf2Z} are the number of thermal pulses during the AGB and
the total  mass lost during the  entire AGB phase, in  solar units. We
find  that the  number  of thermal  pulses  during the  AGB phase  is,
generally  speaking,  slightly  larger   for  the  set  of  metal-poor
evolutionary  sequences.   For  instance,  for  the  $1.5\,  M_{\sun}$
stellar  sequence a  total of  7 thermal  pulses occur  for  the solar
metallicity   white  dwarf  progenitor,   while  for   the  metal-poor
progenitor  this  number  is  10,  thus leading  to  a  more  extended
mass-loss phase.  However,  the total mass lost during  the entire AGB
phase is smaller for the  case of a metal-poor progenitor --- $0.757\,
M_{\sun}$ for  the same  model star ---  than for  a solar-metallicity
progenitor --- $0.795\, M_{\sun}$.

\begin{figure}
\begin{center}
\includegraphics[clip,width=0.9\columnwidth]{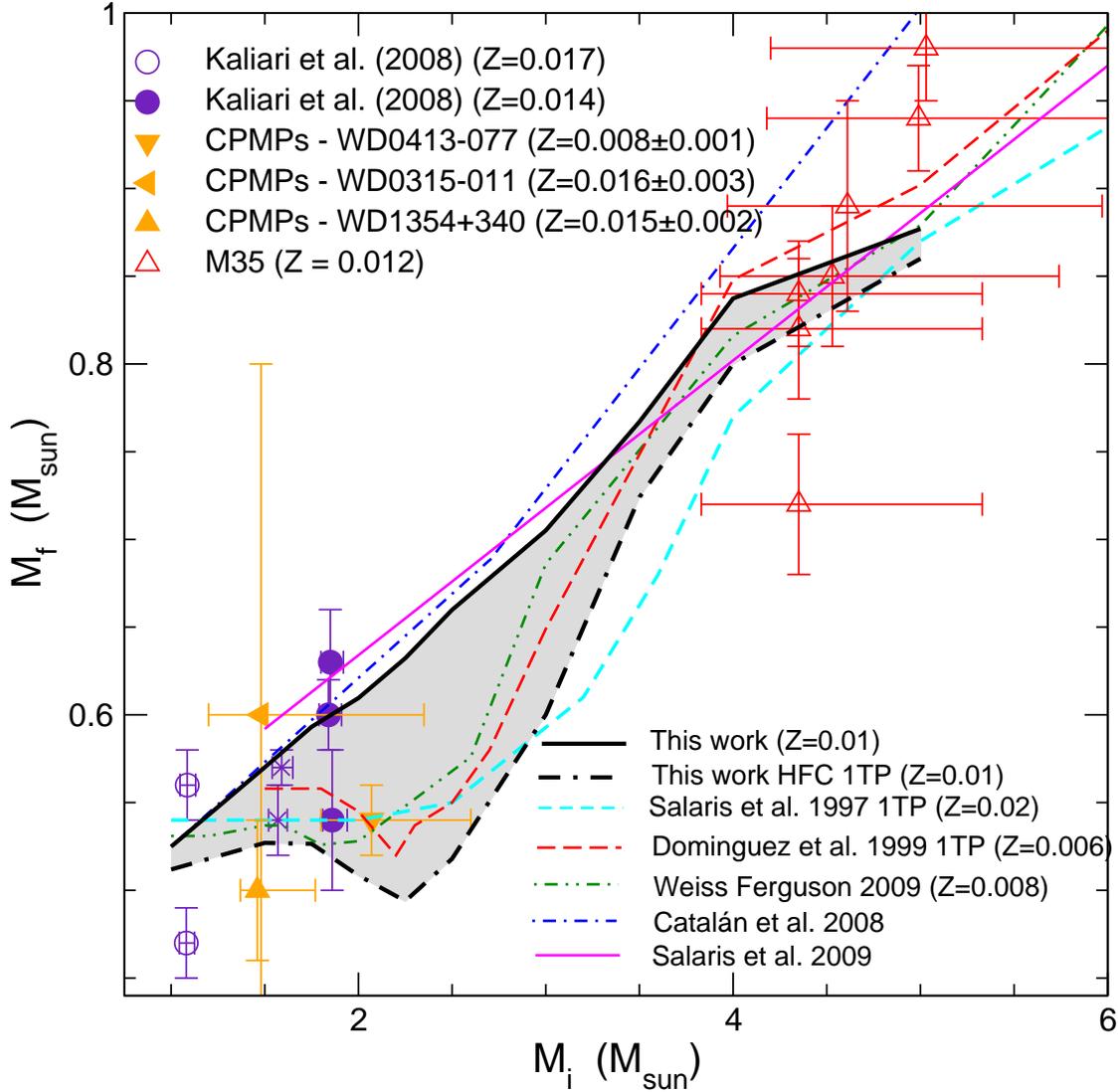}
\end{center}
\caption{Theoretical  initial-to-final  mass  relationship  ---  thick
         solid  line  --- and  mass  of  the  degenerate core  at  the
         beginning   of   the    first   thermal   pulse   ---   thick
         dot-dashed-dashed line  --- obtained  in this work,  both for
         the  case in  which the  solar composition  is  adopted.  The
         initial-to-final mass relationships of Salaris et al.  (1997)
         --- short dashed line --- of Dom\'\i nguez et al.  (1999) ---
         long  dashed line  --- and  of Weiss  \& Ferguson  (2009) ---
         dot-dot-dashed  line --- are  also shown.   The observational
         initial-to-final  mass  relationship   of  Catal\'an  et  al.
         (2008a)  and Salaris  et al.  (2009) are  the  dot-dashed and
         solid  lines, respectively.   See the  online version  of the
         journal for a color version  of this figure and the main text
         for additional details.}
\label{MiMf}
\end{figure}

Perhaps one of  the most interesting results of  our full evolutionary
calculations  is  the  initial-to-final  mass  relationship.  In  Fig.
\ref{MiMf}  we show  our results  for  the case  of solar  composition
($Z=0.01$).  Specifically, in  this figure we show using  a solid line
the  mass   of  the  white   dwarf  resulting  from   our  theoretical
calculations as  a function of the  initial mass at the  ZAMS. For the
sake of  comparison we  also show using  a dot-dashed-dashed  line the
mass of the  hydrogen-free core at the beginning  of the first thermal
pulse as a function of the initial mass of the progenitor star, and we
compare  this relationship  with that  of Salaris  et al.   (1997) ---
short dashed line  --- and Dom\'\i nguez et  al.  (1999) for $Z=0.006$
--- long dashed  line.  These two  initial-to-final mass relationships
were  obtained assuming  that the  mass of  the resulting  white dwarf
corresponds to that of the  hydrogen-free degenerate core at the first
thermal  pulse,   and  consequently  that  the  core   does  not  grow
appreciably  afterwards.    In  addition,   we  also  show   (using  a
dot-dot-dashed  line)  the  recently  obtained  initial-to-final  mass
relationship  of  Weiss \&  Ferguson  (2009)  for  $Z=0.008$. In  this
calculation envelope  overshooting was used during  the TP-AGB, which,
as mentioned  in Sect.  \ref{Input physics}, considerably  reduces the
further  growth   of  the  hydrogen-free  core.   We   also  show  the
observational initial-to-final  mass relationship of  Catal\'an et al.
(2008a) ---  dot-dashed line.  This  relationship is based  on cluster
observations  of  different metallicities,  which  are  close to  that
adopted  in   our  calculations   ($Z=0.01$).   Also  shown   are  the
observational  results  of  Kalirai  et  al.  (2008)  ---  with  their
corresponding error  bars ---  which correspond to  metallicities very
close to the solar metallicity --- filled and open circles --- and the
results for individual  white dwarfs in common proper  motion pairs of
Catal\'an  et al.   (2008b) ---  solid triangles.   The  main sequence
stars of these common proper motion pairs also have solar metallicity.
In  addition,  the results  for  M35 as  quoted  in  Catal\'an et  al.
(2008a), for  which the estimated  metallicity is also close  to solar
(Barrado  y Navascu\'es  et al.  2001)  are also  shown. Finally,  the
semi-empirical  initial-to-final  mass   relationship  of  Salaris  et
al.  (2009)  based  on   open  cluster  observations  is  included  in
Fig. \ref{MiMf}.

It  is to  be noted  the excellent  agreement between  our theoretical
calculations  and the  empirical initial-to-final  mass relationships,
particularly that  of Salaris et al.   (2009).  Note as  well that the
mass of the  hydrogen-free degenerate core at the  first thermal pulse
for all the theoretical sequences  agrees with each other and presents
a minimum  around $\sim 2.0\, M_{\sun}$,  but does not  agree with the
empirical  initial-to-final mass  relationship.   This emphasizes  the
importance  of carefully following  the evolution  of the  star models
from  the main  sequence all  the way  through the  TP-AGB  phase and,
finally, to the  beginning of the white dwarf  cooling track, when the
mass-loss rate  becomes negligible.  In particular, the  growth of the
core mass during the TP-AGB phase is emphasized as a gray area in Fig.
\ref{MiMf}.   The   implication  of   a  proper  computation   of  the
intial-to-final  mass relationship  for the  carbon/oxygen composition
expected in  a white dwarf will  be discussed in  a forthcoming paper.
Note  as  well, that  our  pre-white  dwarf evolutionary  calculations
provide us  with accurate and reliable starting  configurations at the
beginning of  the white  dwarf cooling phase,  as they yield  not only
self-consistent  inner  chemical  profiles,  but also  masses  of  the
hydrogen-rich envelopes,  helium buffers  and core masses  which agree
with the observational results.

\subsection{A global view of the white dwarf cooling phase}

\begin{figure}
\begin{center}
\includegraphics[clip,width=0.9\columnwidth,angle=-90]{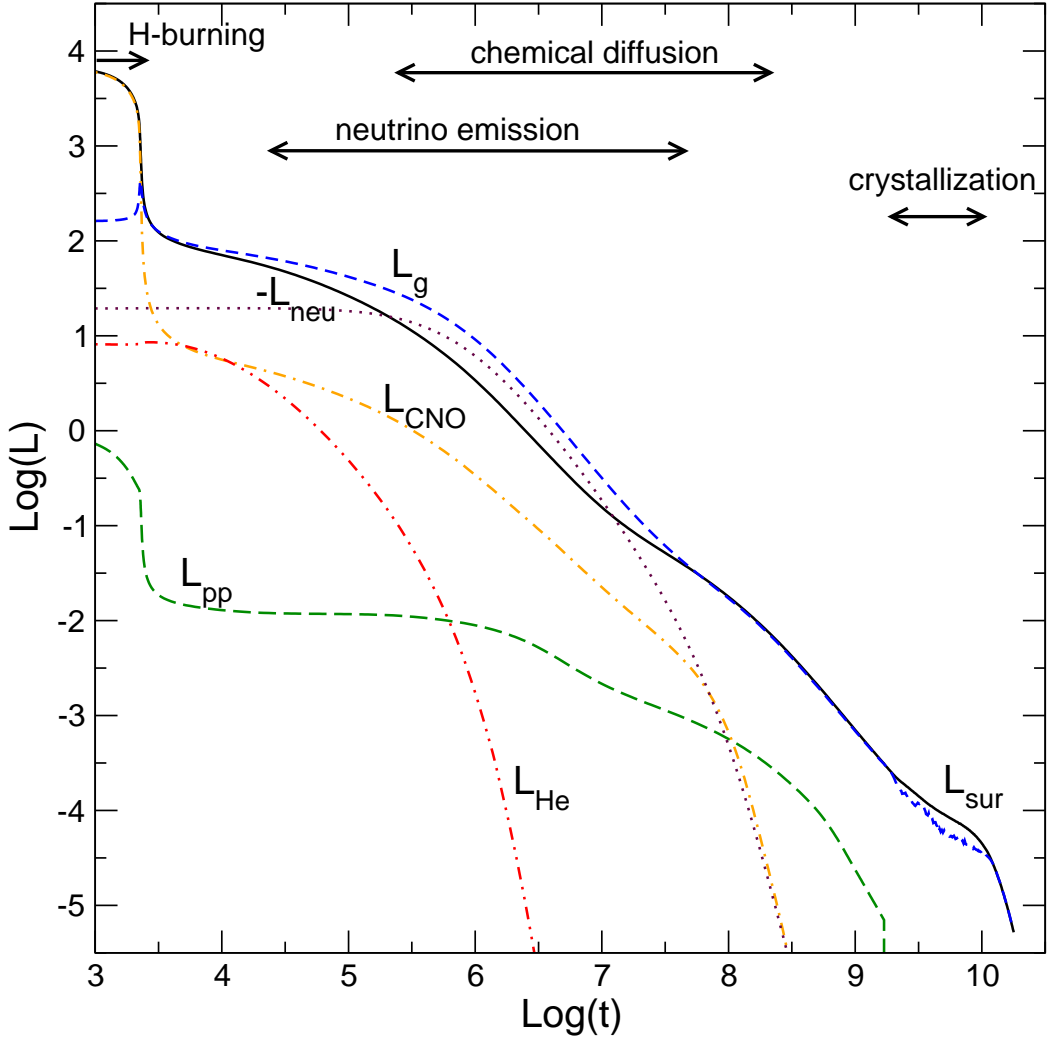}
\end{center}
\caption{Time dependence of the different luminosity contributions for
         our   $0.609\,   M_{\sun}$    white   dwarf   sequence   when
         carbon-oxygen  phase  separation  is  included. We  show  the
         photon   luminosity,   $L_{\rm   sur}$  (solid   line),   the
         luminosities  due  to  nuclear  reactions  ---  proton-proton
         chains, $L_{\rm pp}$ (long dashed line), CNO bicycle, $L_{\rm
         CNO}$  (dot-dashed   line),  helium  burning,   $L_{\rm  He}$
         (dot-dot-dashed line) ---  the neutrino losses, $L_{\rm neu}$
         (dotted line) and the  rate of gravothermal (compression plus
         thermal) energy release, $L_{\rm  g}$ (dashed line).  Time is
         expressed  in years  counted  from the  moment  at which  the
         remnant reaches  $\log T_{\rm eff}=4.87$  at high luminosity.
         The various  physical processes  occuring as the  white dwarf
         cools  down  are  indicated  in the  figure.  The  progenitor
         corresponds to a solar-metallicity $2.0\, M_{\sun}$ star.}
\label{LAge20_A_color3}
\end{figure}

In  Fig.   \ref{LAge20_A_color3}  we  show  the  different  luminosity
contributions   during  the   white  dwarf   cooling  phase,   for  an
archetypical  $0.609\, M_{\sun}$  carbon-oxygen white  dwarf resulting
from a progenitor star of $2.0\, M_{\sun}$ with solar composition. The
very  first phases  of the  cooling  phase are  dominated by  residual
hydrogen burning in the outer layers.  This can be easily seen in Fig.
\ref{LAge20_A_color3},  where   the  different  nuclear  luminosities,
namely the proton-proton  hydrogen-burning luminosity, the CNO bicycle
luminosity and  the helium-burning luminosity are shown  as a function
of  the  cooling  age.   As  can  be seen  in  this  figure,  at  high
luminosities  the largest  contribution  comes from  the CNO  bicycle,
being the proton-proton and  the helium-burning luminosities orders of
magnitude smaller.   This a short-lived  phase (a few  thousand years)
and, thus, given  the long-lived cooling times of  white dwarfs, it is
totally  negligible in  terms  of age.   Nevertheless,  this phase  is
important as  it configures the  final thickness of  the hydrogen-rich
envelope  of  the white  dwarf.   After  this  short-lived phase,  the
nuclear luminosities abruptly decline (at $\log t \simeq 3.6$) and the
release of gravothermal energy  becomes the dominant energy source and
drives  the evolution.   In  passing, we  note  that residual  nuclear
reactions are not  totally extinguished until very late  phases of the
evolution,  in  agreement  with  the  pioneering results  of  Iben  \&
MacDonald (1985, 1986).  In  fact, there are still small contributions
of both  the CNO cycle and  proton-proton chains until  $\log t \simeq
8.3$  and   $\log  t  \simeq  9.0$,   respectively.   Although  almost
negligible for the calculation of the cooling age in the case of white
dwarfs  resulting from  solar metallicity  progenitors,  this residual
nuclear burning becomes relevant at  very late stages for white dwarfs
resulting from low-metallicity progenitors as it lasts for one billion
years, see next  section.  The phase in which  the evolution is driven
by gravitational contraction lasts for about one million years. During
this phase the release of gravothermal energy occurs preferentially in
the  outer partially-  or non-degenerate  layers of  the  white dwarf.
More or less at the same epoch --- that is, at $\log(t)\simeq 5.6$ ---
neutrino losses  become also important.  In particular, at  this epoch
neutrinos are the  dominant energy sink in the  degenerate core of the
white dwarf,  and their associated luminosity becomes  larger than the
optical luminosity.  In fact, during  a relatively long period of time
(from  $\log  t  \sim  5.6$  to  $7.1$)  the  neutrino  luminosity  is
comparable  to the  luminosity associated  to the  gravothermal energy
release.  It  is also  interesting to note  that at  approximately the
same   time,   element   diffusion   is   operating   in   the   outer
partially-degenerate envelope, shaping  the chemical stratification of
the very outer  layers of the white dwarf.  We  will discuss below the
resulting  chemical   stratification.  This  phase   lasts  for  about
$2.2\times 10^8$ years.  At $\log  t \sim 7.8$ the temperatures in the
degenerate core  decrease below the threshold  where neutrino emission
ceases   and,   consequently,   the   neutrino   luminosity   abruptly
drops. During this phase of  the evolution most of the energy released
by the white dwarf has  gravothermal origin, and the white dwarf cools
according to  the classical Mestel's  law (Mestel 1952).   Finally, at
$\log t \simeq 9.2$ crystallization sets in at the center of the white
dwarf and the cooling process slows  down due to the release of latent
heat  and   of  gravitational   energy  due  to   carbon-oxygen  phase
separation.  These  physical processes are  noticeable as a  change in
the  slope  of  the cooling  curve.   Note  as  well that  during  the
crystallization  phase   the  surface   energy  is  larger   than  the
gravothermal luminosity,  a consequence     of these two  additional energy
sources.  This phase lasts for  $\sim 9.4\times 10^9$ years. After this
phase, the  temperature of  the crystallized core  of the  white dwarf
drops below the  Debye temperature and the heat  capacity of the white
dwarf  drops.  Consequently,  the  white dwarfs  enters the  so-called
Debye  cooling phase,  and the  slope of  the cooling  curve increases
again.  This occurs at $\log t \sim 10$.

\subsection{The thickness of the hydrogen envelope}

\begin{figure}
\begin{center}
\includegraphics[clip,width=0.9\columnwidth]{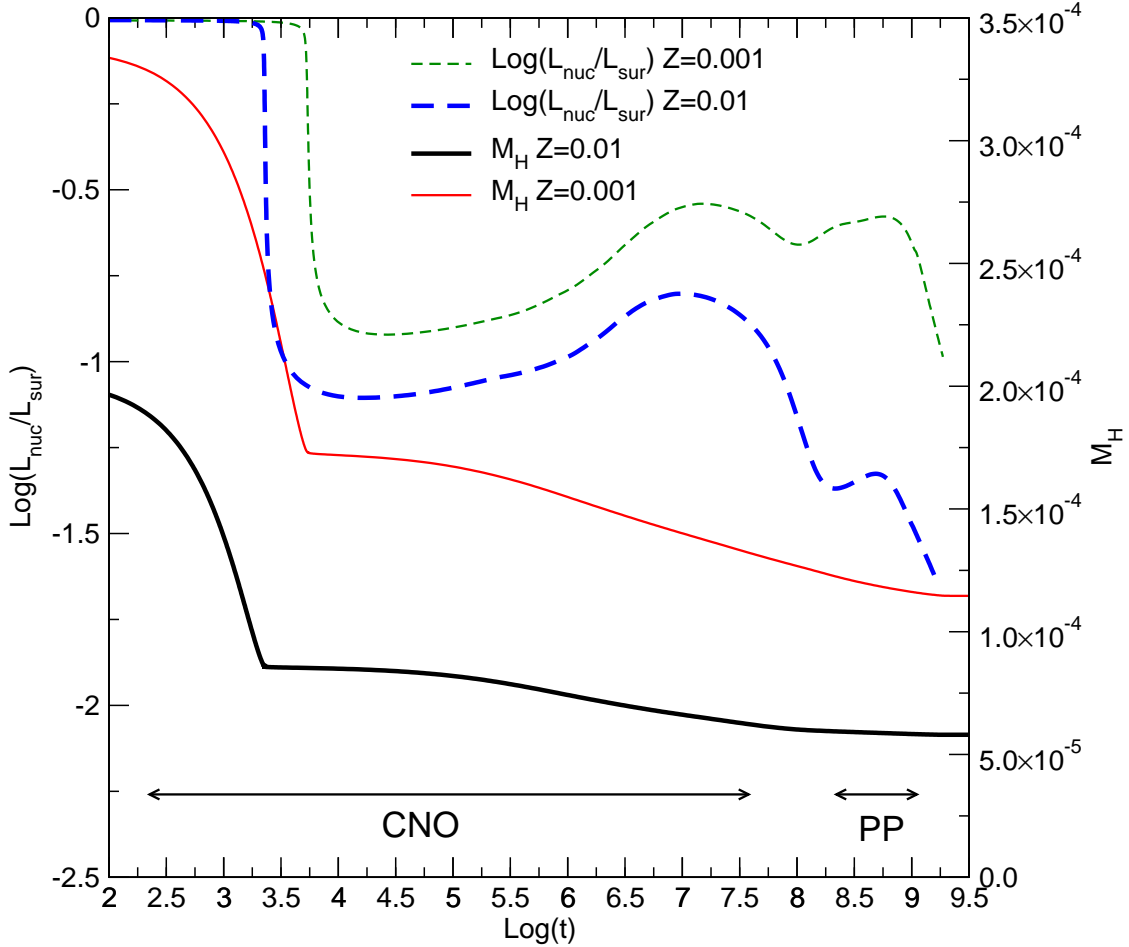}
\end{center}
\caption{Temporal evolution  of the  hydrogen content $M_{\rm  H}$ (in
         solar masses) and the ratio of hydrogen (proton-proton chains
         and CNO bicycle) nuclear burning to surface luminosity, solid
         and dashed lines, respectively. Thick (thin) lines correspond
         to  progenitors  with  $Z=0.01$  ($Z=0.001$). Note  that  the
         hydrogen content  left in the white dwarf,  and therefore the
         nuclear energy  output, are  strongly dependent on  the metal
         content of the progenitor stars. Although a large fraction of
         the hydrogen content is  burnt before the remnant reaches the
         terminal cooling track  at young ages, note that  in the case
         of low  metallicity, residual burning during  the white dwarf
         stage reduces the hydrogen  content considerably. The mass of
         the  white  dwarf corresponding  to  $Z=0.01$ ($Z=0.001$)  is
         $0.609M_{\sun}$ ($0.593M_{\sun}$).}
\label{MasaH_3}
\end{figure}

Fig. \ref{MasaH_3} shows  the temporal evolution of the  masses of the
hydrogen content for two  representative white dwarf cooling tracks of
the two metallicities explored here.   Also shown are the ratio of the
hydrogen-burning luminosities to  the total luminosity. In particular,
the thick lines represent the  evolution of a $0.609\, M_{\sun}$ white
dwarf resulting  from a  solar progenitor, while  the thin  lines show
that of a  $0.593\, M_{\sun}$ white dwarf resulting  from a metal-poor
progenitor.  The solid  lines correspond to the evolution  of the mass
of the hydrogen content, while  the dashed lines show the evolution of
the nuclear  luminosities.  As can be seen,  residual hydrogen burning
is dominant  during the first  evolutionary phases of the  white dwarf
stage. As a  consequence, the mass of these  envelopes decreases for a
period of  time of  $\sim 3\times 10^3$  years, during  which hydrogen
burning supplies  most of the  surface luminosity of the  white dwarf.
However, as soon as the mass of the hydrogen content decreases below a
certain   threshold  ($\sim  8\times   10^{-5}\,  M_{\sun}$   for  the
solar-metallicity star and $\sim 1.7\times 10^{-4}\, M_{\sun}$ for the
metal-poor star)  the pressure  at the bottom  of the envelope  is not
large enough to support further  nuclear reactions, and hence the main
energy source  of the white dwarf  is no longer  nuclear reactions but
gravothermal  energy  release,  and  the hydrogen  content  reaches  a
stationary value.  This is true in  the case of the white dwarf with a
solar  metallicity progenitor, but  for the  white dwarf  remnant that
results from the  lower metallicity progenitor star, it  is clear that
residual hydrogen  burning is  by no means  negligible. In  this case,
note that  hydrogen burning represents  an importante fraction  of the
surface luminosity after  $\approx 10^7$ yr of evolution,  and even at
more advanced  stages ($\approx 10^9$ yr),  this contribution reaches
up to  30 \%. At  this time, the  nuclear energy production  is almost
entirely from  the proton-proton chain,  and the hydrogen  content has
been  reduced  down  to  $\sim  1.1\times  10^{-4}\,  M_{\sun}$.   The
contribution of  hydrogen burning to surface  luminosity increases for
white dwarfs with lower stellar masses.  We would like to emphasize at
this  point the  importance  of computing  self-consistently the  very
first  stages  of  the  white  dwarf evolution,  as  they  provide  an
homogeneous set of white dwarf envelope masses, which as the evolution
proceeds influence the cooling of white dwarfs.

\subsection{The chemical abundances of the envelope}

\begin{figure}[t]
\begin{center}
\includegraphics[clip,width=0.9\columnwidth]{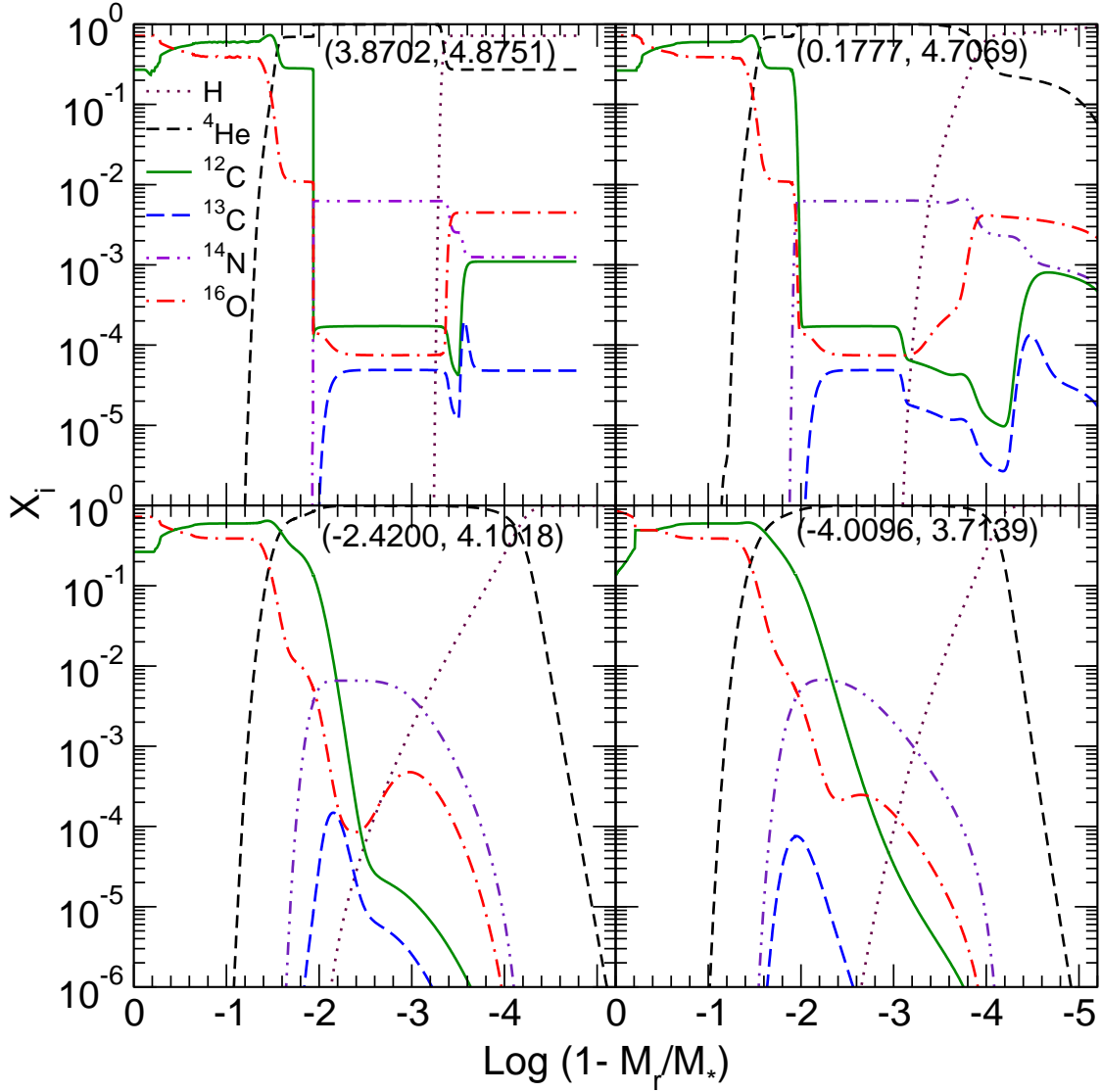}
\end{center}
\caption{Abundance by mass of H, $^4$He, $^{12}$C, $^{13}$C, $^{14}$N,
         $^{16}$O   as  a   function  of   the  outer   mass  fraction
         $\log(1-M_{r}/M_{*})$ for our $0.609 \, M_{\sun}$ white dwarf
         sequence  at  various   selected  evolutionary  stages.   The
         upper-left  panel corresponds  to  the start  of the  cooling
         branch  ($\log  T{\rm  eff}=4.87$  at high  luminosity).  The
         upper-right  panel  shows the  chemical  profiles after  some
         diffusion  has  already taken  place  in  the envelope.   The
         bottom-left panel  shows the situation  at the domain  of the
         pulsating DA  white dwarfs.  Finally,  the bottom-right panel
         shows   the   chemical   abundances   after  the   onset   of
         cyrstallization.    Luminosity  and   effective  temperatures
         ($\log L/L_{\sun}$, $\log T{\rm eff}$) are specified for each
         stage.  The metallicity of the progenitor star is $Z=0.01$.}
\label{PerfilesQuimicosEnvoltura060_3}
\end{figure}

Fig.  \ref{PerfilesQuimicosEnvoltura060_3} shows the chemical profiles
of  the  $0.609\,  M_{\sun}$   white  dwarf  resulting  from  a  solar
metallicity  progenitor  for selected  evolutionary  stages along  the
white dwarf  cooling track.  Each of  the panels is  labelled with the
luminosity  and effective  temperature of  the evolutionary  stage. In
these  panels we  show  the abundance  profiles  of hydrogen,  helium,
carbon, nitrogen  and oxygen in terms  of the outer  mass fraction. As
can  be seen  in  the  upper-left panel,  which  depicts the  chemical
profiles at  the beginning of  the cooling track, the  resulting white
dwarf  has  a  hydrogen-rich  envelope, with  substantial  amounts  of
heavier  elements,  like helium,  carbon,  nitrogen  and oxygen.   The
chemical composition of  this layer is similar to  that of typical AGB
stars that have not experienced third dredge-up episodes, being oxygen
more  abundant  than  carbon,  and  nitrogen  almost  as  abundant  as
carbon. Specifically,  these abundances  are essentially fixed  by the
first dredge-up episode during the  red giant phase. The deeper layers
in  the  helium buffer  zone  show  CNO  abundances that  reflect  the
occurrence of hydrogen burning in prior stages, with nitrogen far more
abundant than  carbon and oxygen.   As the white dwarf  evolves across
the  knee in the  Hertzsprung-Russell diagram,  gravitational settling
and diffusion  become the relevant physical processes  and the heavier
chemical elements  begin to sink appreciably.  This  is illustrated in
the  upper-right panel of  Fig.  \ref{PerfilesQuimicosEnvoltura060_3}.
As  can be seen  in this  panel, at  this stage,  the white  dwarf has
already  developed a  thin  pure hydrogen  envelope  that thickens  as
evolution  proceeds.   Note  that  at  this  evolutionary  stage  some
diffusion and  gravitational settling  has already occurred  in deeper
layers,   and    the   chemical   interfaces    exhibit   less   sharp
discontinuities.   During these stages,  the chemical  composition has
also changed  as a result of nuclear  burning via the CN  cycle at the
base  of the  hydrogen envelope.   With  further cooling  --- see  the
bottom-left  panel of  Fig.   \ref{PerfilesQuimicosEnvoltura060_3} ---
the action of  element diffusion becomes more apparent.   In fact, the
helium-rich buffer increases its size  and both carbon and oxygen sink
towards  deeper and  deeper  regions  of the  white  dwarf. Also,  the
thickness of the hydrogen rich layer increases appreciably, and at the
same  time, the  tail of  the  hydrogen distribution  continues to  
chemically diffusing  inward.  At this stage, which  correponds to the
domain  of the  pulsating DA  white  dwarfs, a  rather thick  hydrogen
envelope has been formed, and below it, a helium-rich and several very
thin layers, which are rich in even heavier elements --- a consequence
of the  high gravity  of the white  dwarf.  Finally,  the bottom-right
panel depicts  the situacion after the onset  of crystallization. Note
the change of carbon and oxygen composition of the core as a result of
crystallization.  This  sequence of figures  emphasizes the importance
of  a proper  treatment of  time-dependent diffusion  processes during
white dwarf  evolution, and the  extent to which the  initial chemical
stratification at the  start of the cooling phase  is altered by these
processes.

\subsection{Convective coupling and crystallization}

\begin{figure}
\begin{center}
\includegraphics[clip,width=0.9\columnwidth,angle=-90]{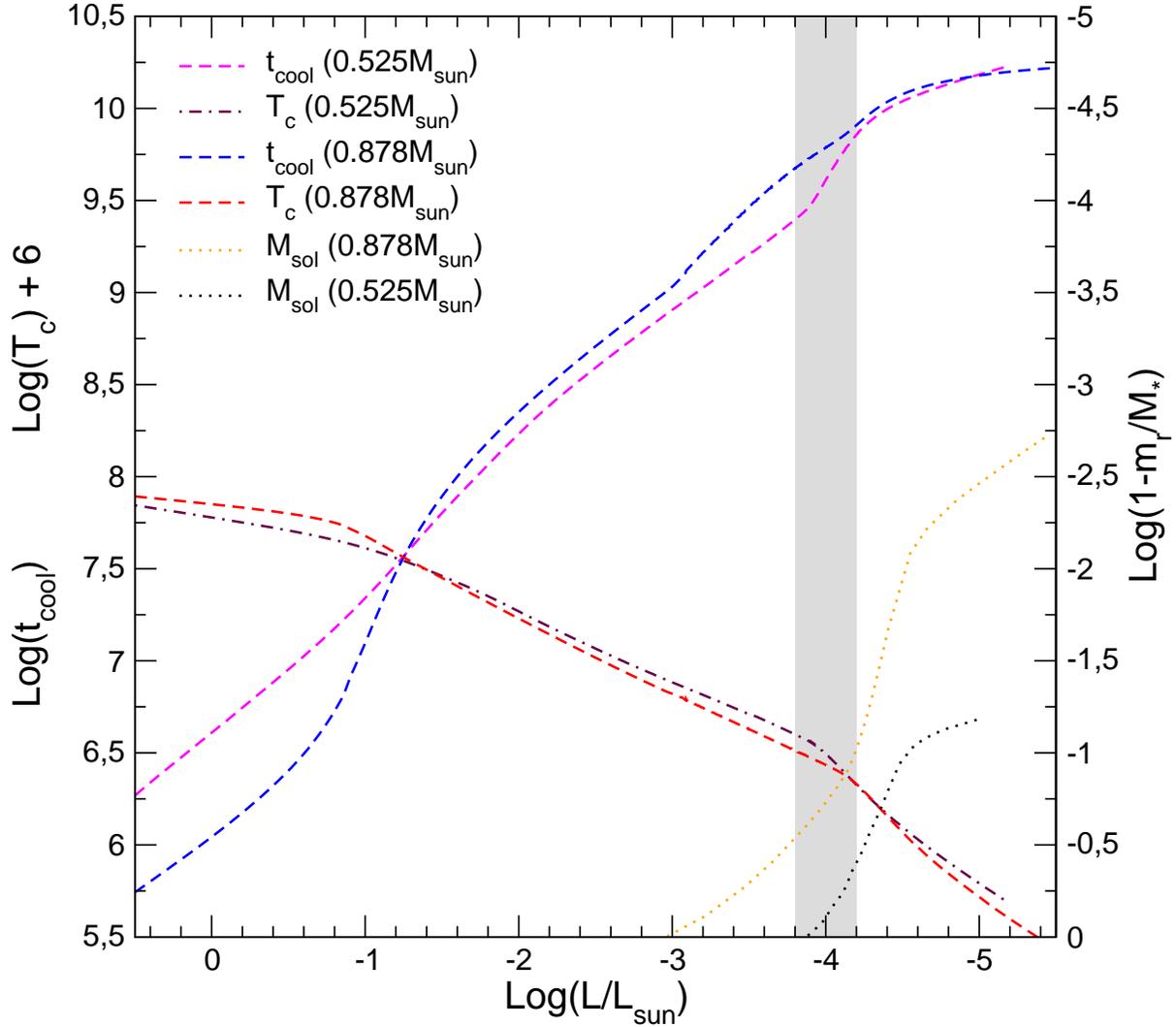}
\end{center}
\caption{Some  evolutionary  properties  corresponding to  our  $0.878
         \,M_{\sun}$ and $0.525 \,M_{\sun}$ white dwarf sequences with
         carbon-oxygen   phase  separation  resulting   from  $Z=0.01$
         progenitors. We  show in terms of the  surface luminosity the
         run  of the  cooling  times (dashed  lines),  of the  central
         temperature   (dot-dashed   lines),    both   read   on   the
         left-hand-side scale, and the  evolution of the growth of the
         crystallized core  (dotted lines) as given by  the outer mass
         fraction on  the right-hand-side  scale. The gray  area marks
         the  occurrence of convective  coupling. The  energy released
         due to convective coupling  and the energy resulting from the
         crystallization process  markedly impact the  cooling curves,
         see text for details.}
\label{convectiveCoupling}
\end{figure}

As discussed in Sect. 3.2, the cooling curve is influenced strongly by
crystallization. However, at this  evolutionary stage the slope of the
cooling curve is  not only dictated by the release  of latent heat and
other  energy sources associated  to crystallization  but also  by the
so-called  convective  coupling.   When  the  envelope  becomes  fully
convective  the  inner  edge  of  the convective  region  reaches  the
boundary  of the  degenerate regions  (Fontaine et  al.   2001).  This
effect is illustrated in Fig.  \ref{convectiveCoupling}, where we show
as a function of the surface luminosity (in solar units) the evolution
of the cooling  times and central temperatures (left  scales), and the
mass of the crystallized white  dwarf core (right scale) for two white
dwarfs, a  low-mass white  dwarf of $0.525  \, M_{\sun}$ and  a rather
massive  white dwarf  with $M=0.878\,  M_{\sun}$, both  resulting from
solar metallicity  progenitors. As  can be seen  there, as  both white
dwarfs cool, there is a  gradual decrease of the central temperatures,
while their  corresponding cooling  times also increase  smoothly.  At
approximately $\log  (L/L_{\sun})\simeq -3.8$ crystallization  sets in
for the  less massive white dwarf,  whereas for the  massive star this
occurs at  $\log (L/L_{\sun})\simeq -3$.   Convective coupling between
the degenerate  core and the partially  degenerate convective envelope
also  occurs  at  low  luminosities.   Since the  inner  edge  of  the
convective envelope reaches  the boundary of the core,  an increase of
the rate of  energy transfer across the outer  opaque envelope occurs,
which  is much  more efficient  than radiative  transfer alone.   As a
consequence,  the relation  between  the central  temperature and  the
surface luminosity experiences a sudden  change of slope, which can be
clearly  seen in  Fig.   \ref{convectiveCoupling}, where  we show  the
region in which convective coupling occurs as a shaded area. Note that
in the case of the more massive white dwarf, convective coupling takes
place   at   luminosities   markedly   lower  than   that   at   which
crystallization starts  in the core. In  fact, more than 90  \% of the
white  dwarf mass  has crystallized  by the  time  convective coupling
occurs in  the $0.878\, M_{\sun}$  white dwarf.  In contrast,  for the
less massive white dwarf, both crystallization and convective coupling
occur  at approximately  the  same stellar  luminosity,  and thus  the
resulting  impact of  these effects  on the  rate of  cooling  is more
noticeable in this case.

\subsection{Cooling times and chemical composition of the core}

One of the most noticeable  features of the white dwarf cooling tracks
presented here  is the  inclusion in a  self-consistent manner  of the
release of gravitational energy due  to phase separation of carbon and
oxygen upon  crystallization.  Previous studies of  this kind (Salaris
et  al.   2000)  included  the  effects of  phase  separation,  but  a
semianalytical approach was used. To highlight the importance of phase
separation upon crystallization we have computed two different sets of
cooling  sequences.  In  the first  of these  cooling  sequences phase
separation of carbon and oxygen  was fully taken into account, whereas
in the  second it was disregarded.   In Table \ref{cosun}  we list for
various luminosities the cooling ages of all our white dwarf sequences
resulting from solar metallicity progenitors, when carbon-oxygen phase
separation is  neglected (top  section), and the  corresponding delays
introduced by  carbon-oxygen phase separation.  We also  show the same
quantities for the case of metal-poor progenitors in Table \ref{comp}.
Clearly, phase separation of  carbon and oxygen introduces significant
delays  at low luminosities,  between 1.0  and 1.8  Gyr.  It  is worth
mentioning  that  at  $\log(L/L_{\sun})=-4.6$, a  luminosity  slightly
smaller than  that of  the observed drop-off  in the disk  white dwarf
luminosity function, $\log(L/L_{\sun})  \simeq -4.5$, phase separation
of carbon  and oxygen  represents a correction  of $\sim 15\%$  to the
total  age,  that  although  not  very  large  it  is  not  negligible
whatsoever if  precise cosmochronology is  to be performed.  At $\log(L/L_{\sun})=-4.0$,
the delays constitute  $20 - 25\%$ of the age for the more massive white dwarfs.
Note that
the  magnitude of  the delays  increases with  the mass  of  the white
dwarf.  For  the case   of  white  dwarfs  resulting from  metal  poor
progenitors  (see  Table  \ref{comp}),   and  for  the  same  fiducial
luminosity the delays introduced by carbon-oxygen phase separation are
slightly larger  for the same  white dwarf mass.  Our  computed delays
are larger than  those obtained by Salaris  et al. (2000).   For instance, for
our   $0.609\,    M_{\sun}$   white   dwarf    cooling   sequence   at
$\log(L/L_{\sun})=-4.6$ we  obtain $\delta  t \simeq 1.38$  Gyr, while
Salaris et al. (2000) at the same luminosity obtain for their $0.61 \,
M_{\sun}$  white dwarf  evolutionary sequence  $\delta t  \simeq 1.00$
Gyr.  This difference stems in  part from the larger carbon abundances
of our  white dwarf model, which  leads to a larger  energy release of
the carbon-oxygen phase separation process, and consequently to larger
time delays.   Indeed, the  chemical profiles used  by Salaris  et al.
(2000) were  those  of Salaris  et  al.   (1997).  The central  carbon
abundance for  the $0.61\, M_{\sun}$  white dwarf is  $X_{\rm C}\simeq
0.25$,  while  for our  $0.609\,  M_{\sun}$  model  we obtain  $X_{\rm
C}\simeq 0.29$.  Hence, the  delays introduced by  carbon-oxygen phase
separation are  correspondingly larger in our model.  A realistic core
composition is crucial  for a proper assessment of  the energy release
from phase separation and its impact on the cooling times.

\begin{figure}
\begin{center}
\includegraphics[clip,width=0.9\columnwidth]{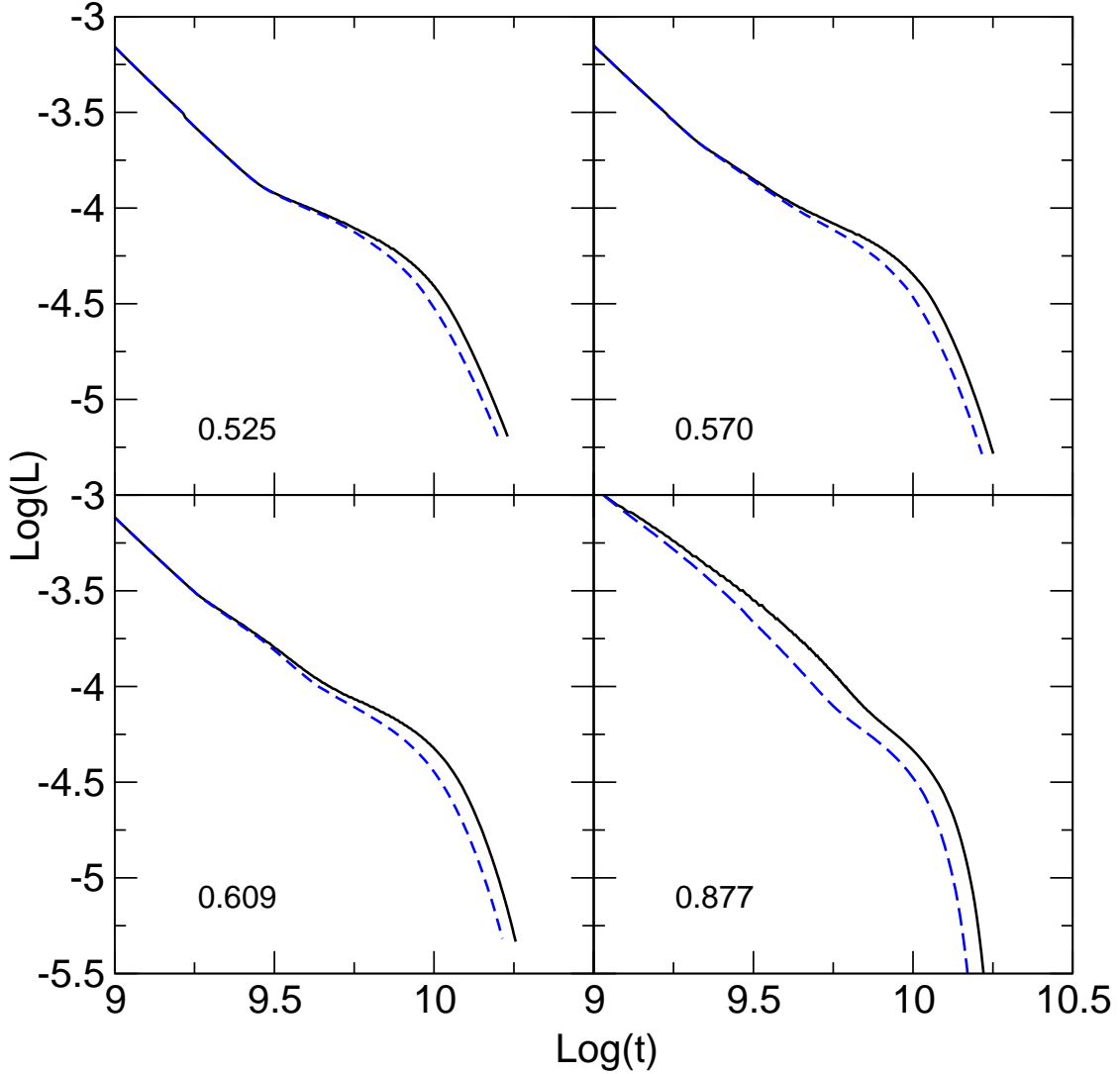}
\end{center}
\caption{Cooling  curves  at  advanced   stages  in  the  white  dwarf
         evolution  for  our sequences  of  masses $0.525\,  M_{\sun}$
         (upper-left  panel), $0.570\, M_{\sun}$  (upper-right panel),
         $0.609M_{\sun}$  (bottom-left   panel),  and  $0.877M_{\sun}$
         (bottom-right panel).  Solid lines correspond to  the case in
         which both latent heat and carbon-oxygen phase separation are
         considered,  while dashed lines  correspond to  the situation
         when  carbon-oxygen   phase  separation  is   neglected.  The
         metallicity of progenitor stars is $Z=0.01$.}
\label{LAge_4grafs_color_3}
\end{figure}

Fig.    \ref{LAge_4grafs_color_3}  shows   the   evolutionary  cooling
sequences of  several selected white dwarfs. Specifically  we show the
luminosity  as a function  of the  cooling age  for white  dwarfs with
solar   metallicity   progenitors   and  masses   $0.525\,   M_{\sun}$
(upper-left panel),  $0.570\, M_{\sun}$ (upper-right  panel), $0.609\,
M_{\sun}$  (bottom-left panel),  and $0.877\,  M_{\sun}$ (bottom-right
panel),  respectively. The figure  emphasizes the  evolutionary stages
where the processes of  convective coupling, cyrstallization and Debye
cooling  take   place.   As  mentioned,  in   low-mass  white  dwarfs,
cyrstallization  and convective  coupling occur  approximately  at the
same  luminosity, $\log(L/L_{\sun})\approx  -4$, thus  resulting  in a
pronounced  impact on  the  cooling  rate.  As  can  be observed,  the
cooling  tracks presented  here have  been computed  down to  very low
luminosities, typically of the  order of $10^{-5}\, L_{\sun}$, or even
smaller.  At  these very low  luminosities the central regions  of the
white dwarf have low enough  temperatures to enter the so-called Debye
cooling  phase. In  this phase  the  specific heat  drops abruptly  as
$T^3$, and as  a consequence the cooling rate  is enhanced.  Thus, the
cooling curve rapidly  drops. The transition to this  phase of cooling
depends  on the  Debye temperature,  $\theta_{\rm D}$,  which  in turn
depends   on    the   density   $(\theta_{\rm   D}\propto\rho^{1/2}$).
Consequently,  more massive white  dwarfs enter  this phase  at larger
temperatures  (luminosities).   In our  most  massive sequence,  rapid
Debye  cooling is  expected to  occur  at the  lowest luminosities  we
computed, as  it is clear from  Fig.  \ref{LAge_4grafs_color_3}, while
for  lower  stellar  masses,  this  phase is  delayed  to  much  lower
luminosities.

\begin{figure}
\begin{center}
\includegraphics[clip,width=0.9\columnwidth]{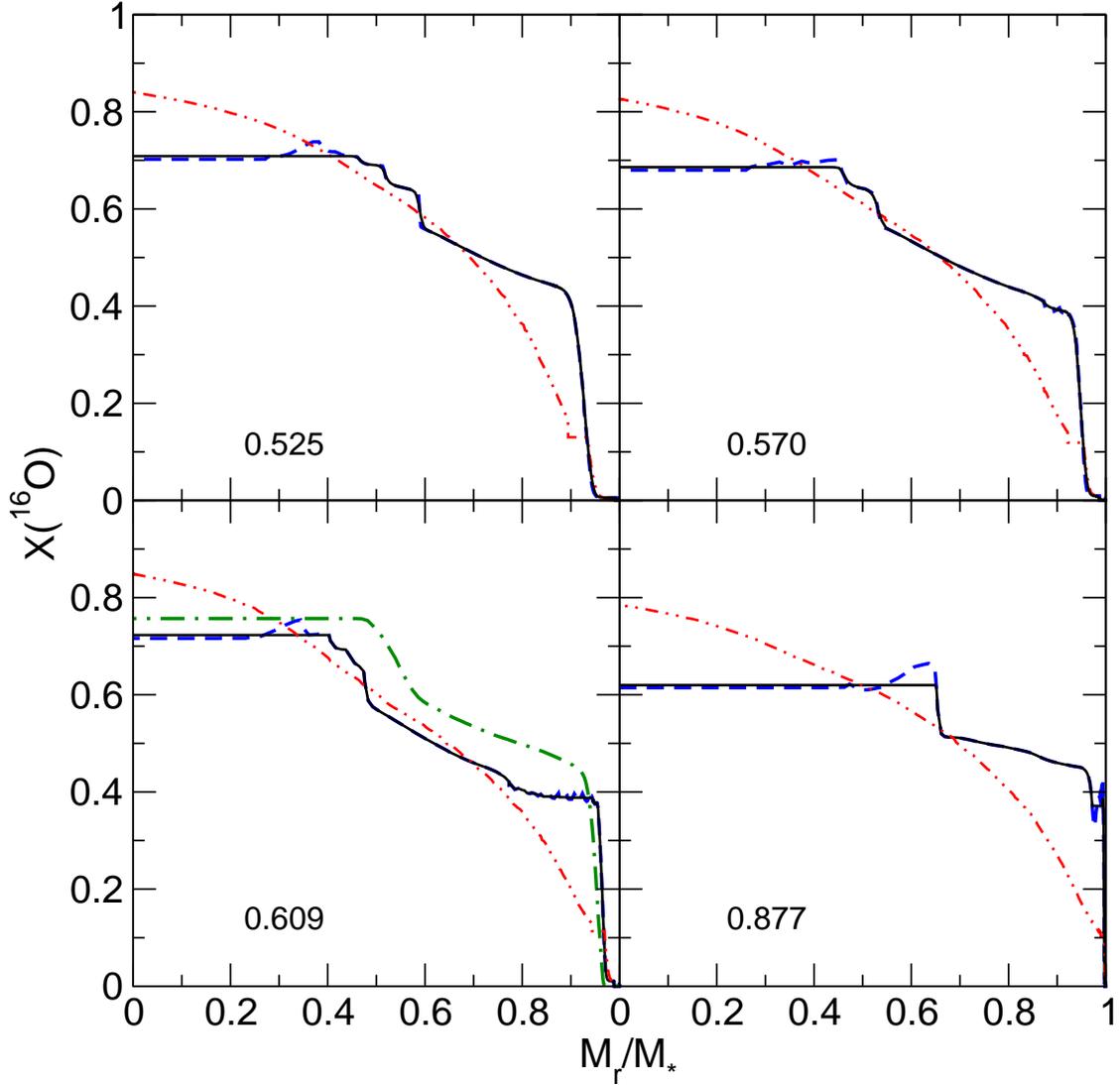}
\end{center}
\caption{Oxygen abundance  (by mass) profiles for  four selected white
         dwarfs  sequences with  stellar masses  (in solar  masses) as
         labeled  in  each panel.   Dashed  lines  show the  abundance
         distribution shortly  after the AGB stage.   Solid lines show
         the  chemical profile after  Rayleigh-Taylor rehomogenization
         has occurred  in the core  and dot-dot-dashed lines  show the
         profiles   after    carbon-oxygen   phase   separation   upon
         crystallization has finished. In the interests of comparison,
         in the bottom-left panel we also show with a dash-dotted line
         the  oxygen  profile of  Salaris  et  al.   (1997) after  the
         rehomogenization phase  of a  white dwarf with  mass $0.609\,
         M_{\sun}$.   Note that  in  all cases,  the initial  chemical
         distribution  in the core  has been  markedly altered  by the
         crystallization process.   The metallicity of  the progenitor
         stars is $Z=0.01$.}
\label{PerfilesQuimicos4Salaris_4}
\end{figure}

Phase  separation in  the  deep  interiors of  white  dwarfs also  has
obvious imprints  in the chemical  profiles of carbon and  oxygen.  To
illustrate this, in  Fig.  \ref{PerfilesQuimicos4Salaris_4} we display
for four  selected white dwarf evolutionary sequences  the oxygen mass
abundance  as  a function  of  the  interior  mass at  three  selected
evolutionary  stages.  Specifically,  we show  the  abundance profiles
shortly after the progenitor star departs from the AGB (dashed lines),
the same profiles  after Rayleigh-Taylor rehomogeneization (Salaris et
al. 1997)  has occurred  (solid lines), and  finally, when  the entire
white  dwarf core  has crystallized  (dot-dot-dashed lines).   For the
sake of  comparison in  the bottom-left panel  of this figure  we also
show  as a  dashed-dotted  line  the profile  obtained  by Salaris  et
al. (1997) for  a white dwarf of $0.61\, M_{\sun}$,  a mass value very
close to that  of this panel.  Note that  the central oxygen abundance
in  the $0.61\, M_{\sun}$  white dwarf  of Salaris  et al.   (1997) is
somewhat higher than that of  our $0.609\, M_{\sun}$ white dwarf. This
is mostly  because we use the  value of the NACRE  compilation for the
$^{12}$C($\alpha,\gamma)^{16}$O  reaction rate  (Angulo et  al.  1999)
which is smaller  than the rate of Caughlan et  al.  (1985) adopted by
Salaris  et  al.  (1997).   However,  another  point  that has  to  be
considered in this comparison is the fact that the white dwarf mass in
Salaris et al.  (1997) is  assessed at the first thermal pulse.  Thus,
for a  given white dwarf mass,  our progenitor stars  are less massive
than those of Salaris et  al. (1997), see our Fig. \ref{MiMf}. Because
of this effect alone, our white dwarf model should be characterized by
a  higher central oxygen abundance  than it  would have  resulted from  a more
massive progenitor  --- see Althaus et  al. (2010).  Thus,  on the one
hand  we expect   a   {\sl lower}  oxygen    abundance   in  our model
because     of     our    adopted     cross     section    for     the
$^{12}$C($\alpha,\gamma)^{16}$O reaction rate,  but on the other hand,
we expect a {\sl higher} oxygen abundance because of the lower initial
mass of our progenitor star. The net effect is that the central oxygen
abundance in our model results somewhat lower than that in the Salaris
et al. (1997) model.  Finally,  the treatment of extra mixing episodes
during core helium burning, which are well known to influence the final carbon oxygen
stratification,   leads   to   some   differences  in   the   expected
composition. In our simulation,  extra-mixing episodes is treated as a
diffusion  processes (Herwig  et al.  1997), while  in Salaris  et al.
(1997) extra-mixing  is   considered  as  a   semiconvective  process.
However, as shown by Straniero et al. (2003) both treatments give rise
to  a  quite  similar   core  chemical  stratification,  and  thus  no
appreciable  difference  in the  central  oxygen  abundance should  be
expected from these treatments.

\subsection{Colors and the blue hook}

\begin{figure}
\begin{center}
\includegraphics[clip,width=0.9\columnwidth]{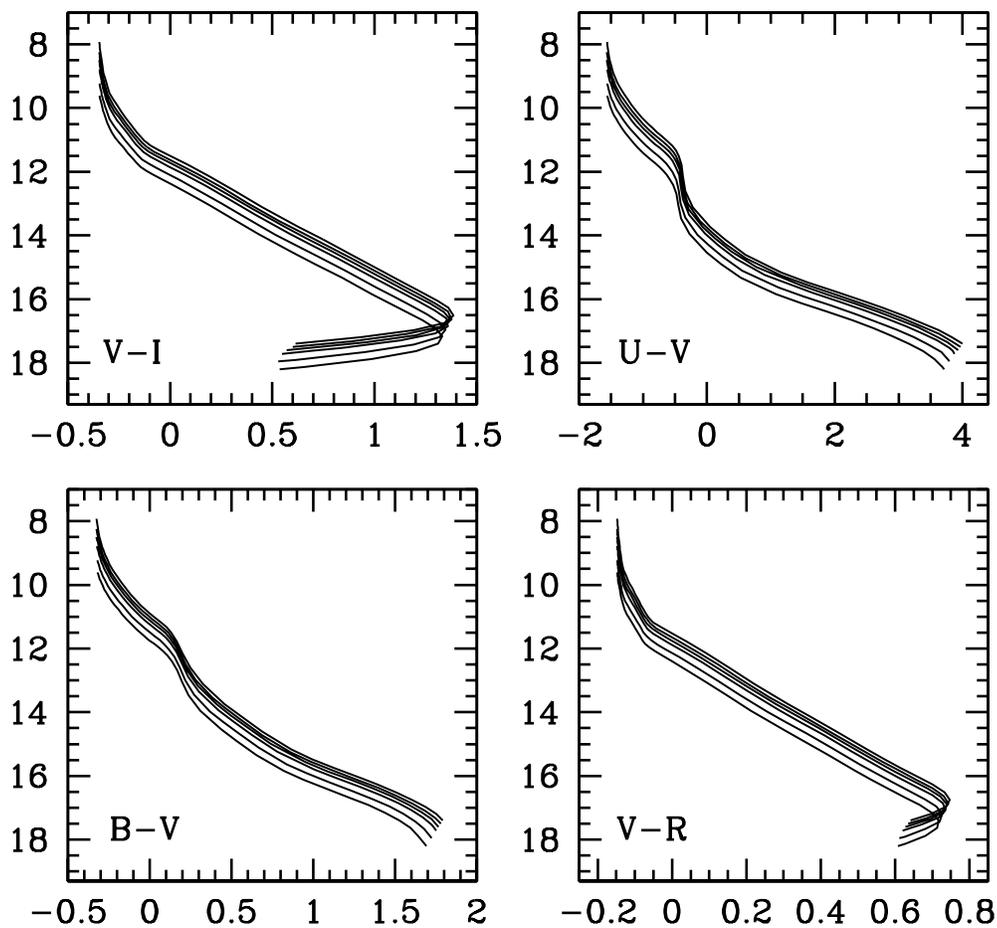}
\end{center}
\caption{Absolute visual  magnitude $M_V$ in terms of  the color index
         $V - I$ (upper-left panel), $U - V$ (upper-right panel), $B -
         V$ (bottom-left panel), and $V - R$ (bottom-right panel), for
         the  complete  evolutionary  tracks  of  our  sequences  with
         $Z=0.01$,  from top to  bottom: $0.525\,  M_{\sun}$, $0.570\,
         M_{\sun}$,  $0.609\, M_{\sun}$, $0.659\,  M_{\sun}$, $0.767\,
         M_{\sun}$, and $0.878\, M_{\sun}$.}
\label{colores2_2}
\end{figure}

The molecular  hydrogen formation  at low effective  temperatures also
affects the evolution of our models in the color-magnitude diagram, as
shown  in  Fig.   \ref{colores2_2},  which  displays the  run  of  the
absolute visual magnitude $M_V$ in  terms of four standard colours: $V
- I$, $U  - V$, $B  - V$,  and $V -  R$.  For the  evolutionary stages
computed  in this  work the  turn to  the blue  at $M_V\approx  17$ is
noticeable for  the $V-I$ and $V-R$ colors.  This effect is due to  the 
 H$_2$-H$_2$ collision-induced absorption over the infrared
spectral regions, which forces stellar flux to emerge at shorter wavelenghts.
Note  that in
this diagram, all  our sequences are expected to  become markedly blue
at $V  - I$ and  $V - R$.  In the $V -  I$ color index,  the turn-off
point occurs between $M_V= 16.5$  and $17.2$, depending on the stellar
mass   value. This   range  corresponds   to   luminosities  between
$\log(L/L_{\sun})= -4.6$ and $-5.0$,  and cooling ages between 11.9 and
15 Gyr for our sequences with carbon oxygen phase separation.

On the  other hand, $U -  V$ and $B -  V$ colors are  sensitive to the
Ly$\alpha$  broadening  by   H-H$_2$  collisions,  which  reduces  the
emergent flux at ultraviolet and blue regions increasing the reddening
of  these colors  in the models cooler than
about $T_{\rm eff}\ = 5000$ K ( $U$ - $V > 1.0$ and $B$ - $V > 0.8$).


\section{Conclusions}
\label{conclusiones}

We have  computed a set  of cooling sequences for  hydrogen-rich white
dwarfs,   which   are    appropriate   for   precision   white   dwarf
cosmochronology.     Our     evolutionary    sequences    have    been
self-consistently evolved from the zero age main sequence, through the
core hydrogen and helium  burning evolutionary phases to the thermally
pulsing asymptotic  giant branch and,  ultimately, to the  white dwarf
stage.   This has  been  done  for white  dwarf  progenitors with  two
different metallicities.  For the  first set of evolutionary sequences
we have adopted solar metallicity.   This allows us to obtain accurate
ages for  white dwarfs in the  local Galactic disk. The  second set of
cooling  sequences  corresponds  to  a  metallicity  typical  of  most
Galactic  globular  clusters,   $Z=0.001$,  thus  allowing  to  obtain
accurate  ages for  metal-poor stellar  systems.  To the  best of  our
knowledge,  this  is the  first  set  of self-consistent  evolutionary
sequences covering different initial masses and metallicities.

Our  main   findings  can  be  summarized  as   follows.   First,  our
evolutionary    sequences    correctly    reproduce    the    observed
initial-to-final mass relationship  of white dwarfs. Specifically, our
evolutionary calculations  are in excellent agreement  with the recent
results  of  Salaris  et  al.   (2009) for  white  dwarfs  with  solar
metallicity  progenitors.  Second, we  corroborate  the importance  of
convective coupling at low luminosity  in the cooling of white dwarfs,
as  originally  suggested  by  Fontaine  et  al.  (2001).   Third,  we
demonstrate  the  importance of  residual  hydrogen  burning in  white
dwarfs resulting from  low-metallicity progenitors. Fourth, we confirm
as  well  the  importance   of  carbon-oxygen  phase  separation  upon
crystallization,  in good  qualitative agreement  with the  results of
Garc\'\i a--Berro et al.  (1988a;  1988b), Segretain et al. (1994) and
Salaris et al. (1997; 2000).  Although the computed delays are smaller
than those previously estimated by  Segretain et al.  (1994), they are
larger than  those computed by Salaris  et al.  (2000), and  are by no
means  negligible if precision  white dwarf  cosmochronology is  to be
done.   However, we  would like  to mention  that these  delays depend
crucially  on  the  previous   evolutionary  history  of  white  dwarf
progenitors    and,    particularly,    on    the    rate    of    the
$^{12}$C$(\alpha,\gamma)^{16}$O  nuclear reaction, as  well as  on the
adopted  treatment  for convective  mixing.   Additionally, since  our
evolutionary  sequences   rely  on  state-of-the-art   non-gray  model
atmospheres,  they reproduce  the  well-known blue  hook  of very  old
hydrogen-rich  white dwarfs  caused  by H$_2$-H$_2$  collision-induced
absorption  (Hansen et  al.  1999).  Finally, we  show  the impact  of
Ly$\alpha$  quasi-molecular opacity  on  the evolution  of cool  white
dwarfs in the color-magnitude diagram.

We  would like  to emphasize  that our  full treatment  of  the entire
evolutionary  history  of  white  dwarfs  has  allowed  us  to  obtain
consistent white dwarf initial configurations. In particular, the mass
of the hydrogen-rich  envelope and of the helium  buffer were obtained
from evolutionary  calculations, instead  of using typical  values and
artificial initial white dwarf  models.  This has implications for the
cooling rates of  old white dwarfs, as the  thicknesses of these outer
layers  control  the cooling  speed  of  such  white dwarfs.   Another
important  issue  which   we  would  like  to  mention   is  that  our
calculations  also yield  self-consistent interior  chemical profiles.
This  also has  relevance  for the  cooling  of white  dwarfs, as  the
release of  latent heat and gravitational energy  due to carbon-oxygen
phase separation upon crystallization crucially depend on the previous
evolutionary   history   of   white   dwarfs.   Also,   the   chemical
stratification of  white dwarf progenitors is  important for correctly
computing the specific  heat of white dwarf interiors.   We would like
to stress as well that the evolutionary tracks of cooling white dwarfs
presented  here has  been  computed with  the  most accurate  physical
inputs and  with a high degree  of detail and  realism. In particular,
our calculations include  nuclear burning at the very  early phases of
white  dwarf evolution  --- which  is important  to determine  the final
thickness   of   the   hydrogen-rich   envelope  ---   diffusion   and
gravitational settling  --- which are important to  shape the profiles
of the  outer layers  --- accurate neutrino  emission rates  --- which
control the cooling at high luminosities --- crystallization and phase
separation of carbon  and oxygen --- which dominate  the cooling times
at low luminosities --- a very detailed equation of state --- which is
important  in all the  evolutionary phases  --- and  improved non-gray
model atmospheres --- which allow for a precise determination of white
dwarf colors  and outer boundary  conditions for the  evolving models.
Finally, we would like to remark that these evolutionary sequences are
important  as well for  the calculation  of self-consistent  models of
pulsating DA  white dwarfs. All in  all, we consider  that our cooling
sequences  represent a  landmark  in the  calculation  of white  dwarf
evolutionary tracks, and they open an new era to precision white dwarf
cosmochronology.  Detailed
tabulations of our evolutionary sequences are available at our
Web site {\tt http://www.fcaglp.unlp.edu.ar/evolgroup}.


\acknowledgements This research was  supported by AGENCIA: Programa de
Modernizaci\'on Tecnol\'ogica  BID 1728/OC-AR, by the  AGAUR, by MCINN
grant AYA2008--04211--C02--01,  by the European Union  FEDER funds and
by PIP  112-200801-00940 grant from CONICET.  LGA  also acknowledges a
PIV  grant of  the AGAUR  of the  Generalitat de  Catalunya.   IR also
acknowledges a FPU grant of the Spanish MEC.


\newpage 

\begin{table*}
\scriptsize
\caption{Initial and final stellar mass (in solar units), total masses
         of  H  and He  left  in the  white  dwarf  (in solar  units),
         main-sequence lifetimes  (in Gyr), mass  lost on the  AGB and
         number of thermal pulses on the AGB for the two metallicities
         studied here.}
\begin{center}
\begin{tabular}{ccccccccccccc}
\hline
\hline
\multicolumn{1}{c}{\,} &
\multicolumn{6}{c}{$Z=0.01$}    &   
\multicolumn{6}{c}{$Z=0.001$}   \\
\cline{2-13}
$M_{\rm ZAMS}$ & $M_{\rm WD}$ & $\log M_{\rm H}$ & $\log M_{\rm He}$ & $t_{\rm MS}$ & $\Delta M_{\rm AGB}$ & $N_{\rm TP}$
               & $M_{\rm WD}$ & $\log M_{\rm H}$ & $\log M_{\rm He}$ & $t_{\rm MS}$ & $\Delta M_{\rm AGB}$ & $N_{\rm TP}$\\
\hline
0.85 & ---   & ---      & ---      & ---   & ---      & --- & 0.505 & $-3.441$ & $-1.567$ & 11.885 & $-0.016$ &  2  \\
1.00 & 0.525 & $-3.586$ & $-1.589$ & 9.040 & $-0.104$ &  5  & 0.553 & $-3.577$ & $-1.635$ &  6.406 & $-0.213$ &  3  \\
1.25 & ---   & ---      & ---      & ---   & ---      & --- & 0.593 & $-3.777$ & $-1.840$ &  2.781 & $-0.499$ &  7  \\
1.50 & 0.570 & $-3.839$ & $-1.703$ & 2.236 & $-0.795$ &  7  & 0.627 & $-4.091$ & $-2.010$ &  1.558 & $-0.757$ & 10  \\
1.75 & 0.593 & $-3.950$ & $-1.851$ & 1.444 & $-1.081$ & 10  & 0.660 & $-4.174$ & $-1.936$ &  1.049 & $-1.012$ & 12  \\
2.00 & 0.609 & $-4.054$ & $-1.826$ & 0.974 & $-1.357$ & 15  & 0.693 & $-4.195$ & $-2.125$ &  0.735 & $-1.272$ & 18  \\
2.25 & 0.632 & $-4.143$ & $-1.957$ & 0.713 & $-1.601$ & 22  & ---   & ---      & ---      &  ---   & ---      & --- \\
2.50 & 0.659 & $-4.244$ & $-2.098$ & 0.543 & $-1.825$ & 21  & ---   & ---      & ---      &  ---   & ---      & --- \\
3.00 & 0.705 & $-4.400$ & $-2.270$ & 0.341 & $-2.279$ & 19  & 0.864 & $-4.860$ & $-2.496$ & 0.279  & $-2.02$  & 34  \\
3.50 & 0.767 & $-4.631$ & $-2.376$ & 0.232 & $-2.688$ & 15  & ---   & ---      & ---      &  ---   & ---      & --- \\
4.00 & 0.837 & $-4.864$ & $-2.575$ & 0.166 & $-3.104$ & 17  & ---   & ---      & ---      &  ---   & ---      & --- \\
5.00 & 0.878 & $-4.930$ & $-2.644$ & 0.099 & $-4.029$ & 12  & ---   & ---      & ---      &  ---   & ---      & --- \\
\hline
\hline
\end{tabular}
\end {center}
\tablecomments {The mass  of the hydrogen and helium  contents is given
                at the  point of maximum effective  temperature at the
                beginning of the white dwarf cooling branch}
\label{table_MiMf2Z}
\end{table*}  

\newpage

\begin{table*}
\scriptsize
\caption{Cooling ages when carbon-oxygen phase separation is neglected
         and  the  accumulated  time  delays  introduced  by  chemical
         fractionation   at  crystallization   for   the  evolutionary
         sequences with progenitors with $Z=0.01$.}
\begin{center}
\begin{tabular}{ccccccccccc}
\hline
\hline
\multicolumn{1}{c}{$-\log(L/L_{\sun})$} &
\multicolumn{10}{c}{$t$ (Gyr)} \\
\hline 
& $0.525\, M_{\sun}$ & $0.570\, M_{\sun}$ & $0.593\, M_{\sun}$ & $0.609\, M_{\sun}$ &$0.632\, M_{\sun}$ 
& $0.659\, M_{\sun}$ & $0.705\, M_{\sun}$ & $0.767\, M_{\sun}$ & $0.837\, M_{\sun}$ &$0.878\, M_{\sun}$ \\ 
\cline{2-11}
 2.0 &  0.17 &  0.17 &  0.17 &  0.17 &  0.17 &  0.17 &  0.18 &  0.20 &  0.21 &  0.22 \\
 3.0 &  0.80 &  0.80 &  0.80 &  0.80 &  0.82 &  0.85 &  0.89 &  0.95 &  1.03 &  1.07 \\
 4.0 &  3.97 &  4.17 &  4.26 &  4.25 &  4.26 &  4.34 &  4.57 &  4.75 &  4.89 &  4.92 \\
 4.2 &  6.57 &  6.79 &  6.85 &  6.85 &  6.86 &  6.95 &  7.10 &  7.02 &  6.80 &  6.62 \\
 4.4 &  8.88 &  9.25 &  9.32 &  9.34 &  9.38 &  9.52 &  9.73 &  9.66 &  9.42 &  9.17 \\
 4.6 & 10.69 & 11.03 & 11.13 & 11.19 & 11.25 & 11.40 & 11.65 & 11.59 & 11.36 & 11.08 \\
 4.8 & 12.40 & 12.71 & 12.77 & 12.81 & 12.84 & 12.96 & 13.18 & 13.04 & 12.71 & 12.39 \\
 5.0 & 14.13 & 14.33 & 14.32 & 14.34 & 14.30 & 14.37 & 14.50 & 14.22 & 13.74 & 13.32 \\
\cline{2-11}
\multicolumn{1}{c}{} &
\multicolumn{10}{c}{$\delta t$ (Gyr)}   \\
\cline{2-11}
 3.0 &  0.00 &  0.00 &  0.00 &  0.00 &  0.00 &  0.00 &  0.00 &  0.00 &  0.00 &  0.01 \\
 4.0 &  0.10 &  0.15 &  0.16 &  0.23 &  0.27 &  0.37 &  0.67 &  0.97 &  1.14 &  1.20 \\
 4.2 &  0.64 &  0.87 &  0.88 &  0.97 &  1.00 &  1.12 &  1.39 &  1.47 &  1.48 &  1.49 \\
 4.4 &  1.03 &  1.26 &  1.28 &  1.30 &  1.33 &  1.45 &  1.66 &  1.72 &  1.68 &  1.67 \\
 4.6 &  1.15 &  1.33 &  1.37 &  1.38 &  1.40 &  1.51 &  1.72 &  1.76 &  1.72 &  1.72 \\
\hline
\hline
\end{tabular}
\end{center}
\label{cosun}
\end{table*}

\newpage

\begin{table*}
\scriptsize
\caption{Same as Table 2 but for $Z=0.001$.}
\begin{center}
\begin{tabular}{cccccccc}
\hline
\hline
\multicolumn{1}{c}{$-\log(L/L_{\sun})$} & 
\multicolumn{7}{c}{$t$ (Gyr)} \\
\hline
& $0.505\, M_{\sun}$ & $0.553\, M_{\sun}$ & $0.593\, M_{\sun}$ & $0.627\, M_{\sun}$ & $0.660\, M_{\sun}$ 
& $0.693\, M_{\sun}$  & $0.864\, M_{\sun}$  \\
\cline{2-8}
\cline{2-8}
2.0 &  0.24 &  0.22 &  0.19 &  0.17 &  0.18 &  0.19 & 0.22 \\
3.0 &  1.01 &  1.03 &  0.91 &  0.83 &  0.86 &  0.89 & 1.06 \\
4.0 &  4.15 &  4.49 &  4.50 &  4.40 &  4.48 &  4.59 & 4.98  \\
4.2 &  6.74 &  7.15 &  7.11 &  7.01 &  7.13 &  7.21 & 6.84 \\
4.4 &  9.01 &  9.57 &  9.61 &  9.54 &  9.73 &  9.84 & 9.42 \\
4.6 & 10.81 & 11.39 & 11.43 & 11.43 & 11.63 & 11.74 & 11.37 \\
4.8 & 12.56 & 13.10 & 13.08 & 13.04 & 13.22 & 13.30 & 12.67\\
5.0 & 14.33 & 14.78 & 14.64 & 14.53 & 14.66 & 14.64 & 13.66 \\
\cline{2-8}
\multicolumn{1}{c}{} &
\multicolumn{7}{c}{$\delta t$ (Gyr)}   \\
\cline{2-8}
3.0 &  0.00 &  0.00 &  0.00 &  0.00 &  0.00 &  0.00 & 0.00 \\
4.0 &  0.08 &  0.17 &  0.16 &  0.25 &  0.39 &  0.49 & 1.14\\
4.2 &  0.49 &  0.80 &  0.94 &  1.14 &  1.24 &  1.27 & 1.44 \\
4.4 &  0.90 &  1.18 &  1.34 &  1.49 &  1.55 &  1.57 & 1.63 \\
4.6 &  1.04 &  1.28 &  1.42 &  1.58 &  1.61 &  1.64 & 1.70 \\
\hline
\hline
\end{tabular}
\end{center}
\label{comp}
\end{table*}

\begin{thebibliography}{}

\bibitem[Alexander  \& Ferguson(1994)]{1994ApJ...437..879A} Alexander,
        D.~R., \& Ferguson, J.~W.\ 1994, \apj, 437, 879
\bibitem[Althaus  et  al.(2003)]{2003A&A...404..593A} Althaus,  L.~G.,
        Serenelli,  A.~M., C{\'o}rsico,  A.~H.,  \& Montgomery,  M.~H.
        2003, \aap, 404, 593
\bibitem[Althaus  et  al.(2005)]{2005A&A...435..631A} Althaus,  L.~G.,
        Serenelli, A.~M.,  Panei, J.~A., C{\'o}rsico,  A.~H., Garc\'\i 
        a--Berro, E., \& Sc\'occola, C.~G. 2005a, \aap, 435, 631
\bibitem[Althaus   et   al.(2005b)]{AE05B}   Althaus,  L.~G.,   Miller
        Bertolami,  M.  M., C{\'o}rsico,  A.~H., Garc{\'{\i}}a--Berro,  
        E., \& Gil-Pons, P. 2005b, \aap, 440, L1
\bibitem[Althaus  et  al.(2007)]{2007A&A...465..249A} Althaus,  L.~G.,
        Garc{\'{\i}}a--Berro,  E., Isern,  J., C{\'o}rsico,  A.~H., \&
        Rohrmann, R.~D.\ 2007, \aap, 465, 249
\bibitem[Althaus  et al.(2009a)]{2009A&A...494.1021A} Althaus,  L.~G.,
        C{\'o}rsico,  A.~H.,   Torres,  S.,  \&  Garc{\'{\i}}a--Berro,
        E. 2009a, \aap, 494, 1021
\bibitem[Althaus  et al.(2009b)]{2009ApJ...693L..23A}  Althaus, L.~G.,
        Garc{\'{\i}}a--Berro,    E.,   C{\'o}rsico,    A.~H.,   Miller
        Bertolami, M.~M., \& Romero, A.~D. 2009b, \apjl, 693, L23
\bibitem[Althaus  et al.(2009c)]{2009A&A...502..207A}  Althaus, L.~G.,
        Panei,  J.~A., Romero,  A.~D.,  Rohrmann, R.~D.,  C{\'o}rsico,
        A.~H., Garc{\'{\i}}a--Berro, E.,  \& Miller  Bertolami, M.~M.\
        2009c, \aap, 502, 207
\bibitem[Althaus  et al.(2009c)]{2009ApJ...704.1605A}  Althaus, L.~G.,
        Panei,  J.~A., Miller Bertolami,  M.~M., Garc{\'{\i}}a--Berro,
        E.,  C{\'o}rsico,  A.~H.,  Romero,  A.~D., Kepler,  S.~O.,  \&
        Rohrmann, R.~D.\ 2009d, \apj, 704, 1605
\bibitem[Althaus  et al.  (2010)]{AE10}  Althaus, L.~G.,  C{\'o}rsico,
        A.~H.,   Bischoff-Kim,  A.,   Romero,  A.   D.,   Renedo,  I.,
        Garc{\'{\i}}a--Berro,  E., \&  Miller Bertolami,  M.  M., 2010
        \apj, submitted
\bibitem[Angulo et al.(1999)]{1999NuPhA.656....3A} Angulo, C., et al.
        1999, Nuclear Physics A, 656, 3
\bibitem[Barrado  y  Navascu{\'e}s et  al.(2001)]{2001ApJ...546.1006B}
        Barrado y Navascu{\'e}s, D.,  Stauffer, J.~R., Bouvier, J., \&
        Mart{\'{\i}}n, E.~L.\ 2001, \apj, 546, 1006
\bibitem[Benvenuto  \&  Althaus(1999)]{1999MNRAS.303...30B} Benvenuto,
        O.~G., \& Althaus, L.~G.\ 1999, \mnras, 303, 30
\bibitem[Benvenuto   et   al.(2004)]{2004PhRvD..69h2002B}   Benvenuto,
        O.~G., Garc{\'{\i}}a--Berro, E., \& Isern, J.\ 2004, \prd, 69,
        082002
\bibitem[Bergeron  et  al.(1997)]{1997ApJS..108..339B}  Bergeron,  P.,
        Ruiz, M.~T., \& Leggett, S.~K.\ 1997, \apjs, 108, 339
\bibitem[Burgers(1969)]{1969fecg.book.....B}   Burgers,  J.~M.   1969,
        {\sl  ``Flow  Equations  for  Composite  Gases''},  New  York:
        Academic Press
\bibitem[Casagrande et al.(2007)]{2007MNRAS.382.1516C} Casagrande, L.,
        Flynn, C., Portinari,  L., Girardi, L., \& Jimenez,  R.\ 2007, 
        \mnras, 382, 1516
\bibitem[Cassisi   et  al.(2007)]{2007ApJ...661.1094C}   Cassisi,  S.,
        Potekhin,  A.~Y., Pietrinferni, A.,  Catelan, M.,  \& Salaris,
        M. 2007, \apj, 661, 1094
\bibitem[Catal{\'a}n  et al.(2008a)]{2008MNRAS.387.1693C} Catal{\'a}n,
        S., Isern, J., Garc{\'{\i}}a--Berro,  E., \& Ribas, I.\ 2008b,
        \mnras, 387, 1693
\bibitem[Catal{\'a}n  et  al.(2008)]{2008A&A...477..213C} Catal{\'a}n,
        S.,  Isern, J., Garc{\'{\i}}a--Berro,  E., Ribas,  I., Allende
        Prieto, C., \& Bonanos, A.~Z.\ 2008, \aap, 477, 213
\bibitem[Caughlan   \&   Fowler(1985)]{1985ADNDT..32..197C}  Caughlan,
        G.  R., Fowler, W.  A., Harris,  M. J.,  \& Zimmermann,  B. A.
        1985, Atomic Data \& Nucl. Data Tables, 32, 197
\bibitem[C{\'o}rsico  et  al.(2001)]{2001NewA....6..197C} C{\'o}rsico,
        A.~H.,  Benvenuto,  O.~G.,   Althaus,  L.~G.,  Isern,  J.,  \&
        Garc{\'{\i}}a-Berro, E. 2001, New Astronomy, 6, 197
\bibitem[Diaz-Pinto et al.(1994)]{1994A&A...282...86D} D\'\i az-Pinto,
        A.,  Garc\'\i  a--Berro,  E.,   Hernanz,  M.,  Isern,  J.,  \&
        Mochkovitch, R.  1994, \aap, 282, 86
\bibitem[Dominguez  et al.(1999)]{1999ApJ...524..226D}  Dom\'\i nguez,
        I., Chieffi,  A., Limongi, M.,  \& Straniero, O.\  1999, \apj,
        524, 226
\bibitem[Fontaine  et  al.(2001)]{2001PASP..113..409F}  Fontaine,  G.,
        Brassard, P., \& Bergeron, P.\ 2001, \pasp, 113, 409
\bibitem[Fontaine  \&  Brassard(2008)]{2008PASP..120.1043F}  Fontaine,
        G., \& Brassard, P. 2008, \pasp, 120, 1043
\bibitem[Flynn(2004)]{2004PASA...21..126F}     Flynn,     C.\    2004,
        Pub. Astron.  Soc. Australia, 21, 126
\bibitem[Garcia-Berro  et   al.(1988a)]{1988A&A...193..141G}  Garc\'\i
        a--Berro,  E., Hernanz,  M.,  Mochkovitch, R.,  \& Isern,  J.\
        1988a, \aap, 193, 141
\bibitem[Garcia-Berro  et   al.(1988b)]{1988Natur.333..642G}  Garc\'\i
        a--Berro,  E., Hernanz,  M.,  Isern, J.,  \& Mochkovitch,  R.\
        1988b, \nat, 333, 642
\bibitem[Garcia-Berro   et   al.(1995)]{1995MNRAS.277..801G}  Garc\'\i
        a--Berro,  E., Hernanz,  M.,  Isern, J.,  \& Mochkovitch,  R.\
        1995, \mnras, 277, 801
\bibitem[Garc{\'{\i}}a-Berro     et    al.(1999)]{1999MNRAS.302..173G}
        Garc\'\i a--Berro, E., Torres,  S., Isern, J., \& Burkert, A.\
        1999, \mnras, 302, 173
\bibitem[Grevesse \&  Sauval(1998)]{1998SSRv...85..161G} Grevesse, N.,
        \& Sauval, A.~J.\ 1998, Space Science Reviews, 85, 161
\bibitem[Haft  et al.(1994)]{1994ApJ...425..222H}  Haft,  M., Raffelt,
        G., \& Weiss, A. 1994, \apj, 425, 222
\bibitem[Hansen(1998)]{1998Natur.394..860H}  Hansen,  B.~M.~S.\  1998,
        \nat, 394, 860
\bibitem[Hansen(1999)]{1999ApJ...520..680H}  Hansen,  B.~M.~S.\  1999,
        \apj, 520, 680
\bibitem[Hansen    \&    Liebert(2003)]{2003ARA&A..41..465H}   Hansen,
        B.~M.~S., \& Liebert, J.\ 2003, \araa, 41, 465
\bibitem[Hansen  et al.(2002)]{2002ApJ...574L.155H}  Hansen, B.~M.~S.,
        et al.\ 2002, \apjl, 574, L155
\bibitem[Hansen  et al.(2007)]{2007ApJ...671..380H}  Hansen, B.~M.~S.,
        et al. 2007, \apj, 671, 380
\bibitem[Herwig   et   al.(1997)]{1997A&A...324L..81H}   Herwig,   F.,
        Bl\"ocker,  T., Sch\"onberner, D.,  \& El  Eid, M.\  1997, \aap,
        324, L81
\bibitem[Iben \&  MacDonald(1985)]{1985ApJ...296..540I} Iben, I., Jr.,
        \& MacDonald, J.\ 1985, \apj, 296, 540
\bibitem[Iben \&  MacDonald(1986)]{1986ApJ...301..164I} Iben, I., Jr.,
        \& MacDonald, J.\ 1986, \apj, 301, 164
\bibitem[Iglesias   \&   Rogers(1996)]{1996ApJ...464..943I}  Iglesias,
        C.~A., \& Rogers, F.~J. 1996, \apj, 464, 943
\bibitem[Isern et  al.(1992)]{1992ApJ...392L..23I} Isern, J., Hernanz,
        M., \& Garcia-Berro, E.\ 1992, \apjl, 392, L23
\bibitem[Isern    et   al.(1997)]{1997ApJ...485..308I}    Isern,   J.,
        Mochkovitch, R., Garcia-Berro, E., \& Hernanz, M.\ 1997, \apj,
        485, 308
\bibitem[Isern et  al.(1998)]{1998ApJ...503..239I} Isern, J., Garc\'\i
        a--Berro,  E.,  Hernanz,   M.,  Mochkovitch,  R.,  \&  Torres,
        S. 1998, \apj, 03, 239
\bibitem[Isern et  al.(2000)]{2000ApJ...528..397I} Isern, J., Garc\'\i
        a--Berro, E.,  Hernanz, M., \& Chabrier, G.\  2000, \apj, 528,
        397
\bibitem[Isern    et   al.(2008)]{2008ApJ...682L.109I}    Isern,   J.,
        Garc{\'{\i}}a--Berro,  E.,  Torres,  S., \&  Catal{\'a}n,  S.\
        2008, \apjl, 682, L109
\bibitem[Itoh  et al.(1996)]{1996ApJS..102..411I}  Itoh,  N., Hayashi,
        H., Nishikawa, A., \& Kohyama, Y. 1996, \apjs, 102, 411
\bibitem[Kalirai  et  al.(2001)]{2001AJ....122.3239K} Kalirai,  J.~S.,
        Ventura,  P., Richer, H.~B.,  Fahlman, G.~G.,  Durrell, P.~R.,
        D'Antona, F., \& Marconi, G.\ 2001, \aj, 122, 3239
\bibitem[Kalirai  et  al.(2008)]{2008ApJ...676..594K} Kalirai,  J.~S.,
        Hansen, B.~M.~S., Kelson,  D.~D., Reitzel, D.~B., Rich, R.~M.,
        \& Richer, H.~B.\ 2008, \apj, 676, 594
\bibitem[Ko{\l}os  \& Wolniewicz(1965)]{1965JChPh..43.2429K} Ko{\l}os,
        W., \& Wolniewicz, L.\ 1965, \jcp, 43, 2429
\bibitem[Kowalski   \&   Saumon(2006)]{2006ApJ...651L.137K}  Kowalski,
        P.~M., \& Saumon, D.\ 2006, \apjl, 651, L137
\bibitem[]{} Kulander, K. C., \& Guest, M. F. 1979, J. Phys. B 12, L501
\bibitem[Lugaro et al.(2003)]{2003ApJ...586.1305L} Lugaro, M., Herwig,
        F.,  Lattanzio, J.~C.,  Gallino, R.,  \& Straniero,  O.\ 2003,
        \apj, 586, 1305
\bibitem[Magni \& Mazzitelli(1979)]{1979A&A....72..134M} Magni, G., \&
        Mazzitelli, I.\ 1979, \aap, 72, 134
\bibitem[Marigo(2002)]{2002A&A...387..507M}  Marigo,  P.\ 2002,  \aap,
        387, 507
\bibitem[Marigo \&  Aringer(2009)]{2009A&A...508.1539M} Marigo, P., \&
        Aringer, B.\ 2009, \aap, 508, 1539
\bibitem[Mestel(1952)]{1952MNRAS.112..583M} Mestel,  L. 1952, \mnras,
        112, 583
\bibitem[Miller  Bertolami   \&  Althaus  (2006)]{2006A&A...454..845M}
        Miller Bertolami,  M.~M., \&  Althaus, L.~G. 2006,  \aap, 454,
        845
\bibitem[Miller  Bertolami  et al.(2008)]{2008A&A...491..253M}  Miller
        Bertolami,  M.~M.,  Althaus,  L.~G.,  Unglaub, K.,  \&  Weiss,
        A. 2008, \aap, 491, 253
\bibitem[Petsalakis   et  al.(1988)]{1988JChPh..88.7633P}  Petsalakis,
        I.~D.,  Theodorakopoulos,  G.,  Wright,  J.~S.,  \&  Hamilton,
        I.~P.\ 1988, \jcp, 88, 7633
\bibitem[Prada  Moroni \&  Straniero(2002)]{2002ApJ...581..585P} Prada
        Moroni, P.~G., \& Straniero, O.\ 2002, \apj, 581, 585
\bibitem[Prada  Moroni \&  Straniero(2007)]{2007A&A...466.1043P} Prada
        Moroni, P.~G., \& Straniero, O.\ 2007, \aap, 466, 1043
\bibitem[]{} Roach, A. C., \& Kuntz, P. J. 1986, J. Chem. Phys. 84, 822
\bibitem[Rohrmann et  al.(2002)]{2002MNRAS.335..499R} Rohrmann, R.~D.,
        Serenelli,  A.~M., Althaus, L.~G.,  \& Benvenuto,  O.~G. 2002,
        \mnras, 335, 499
\bibitem[Salaris   et  al.(1997)]{1997ApJ...486..413S}   Salaris,  M.,
        Dom\'\i nguez, I., Garc\'\i  a--Berro, E., Hernanz, M., Isern,
        J., \& Mochkovitch, R.\ 1997, \apj, 486, 413
\bibitem[Salaris   et  al.(2000)]{2000ApJ...544.1036S}   Salaris,  M.,
        Garc{\'{\i}}a--Berro, E.,  Hernanz, M., Isern,  J., \& Saumon,
        D.\ 2000, \apj, 544, 1036
\bibitem[Salaris   et  al.(2009)]{2009ApJ...692.1013S}   Salaris,  M.,
        Serenelli, A., Weiss, A., \& Miller Bertolami, M.\ 2009, \apj,
        692, 1013
\bibitem[Schr{\"o}der       \&       Cuntz(2005)]{2005ApJ...630L..73S}
        Schr{\"o}der, K.-P., \& Cuntz, M.\ 2005, \apjl, 630, L73
\bibitem[Segretain  \& Chabrier(1993)]{1993A&A...271L..13S} Segretain,
        L., \& Chabrier, G.\ 1993, \aap, 271, L13
\bibitem[Segretain  et al.(1994)]{1994ApJ...434..641S}  Segretain, L.,
        Chabrier,  G., Hernanz,  M., Garcia-Berro,  E., Isern,  J., \&
        Mochkovitch, R.\ 1994, \apj, 434, 641
\bibitem[Straniero  et al.(2003)]{2003ApJ...583..878S}  Straniero, O.,
        Dom{\'{\i}}nguez,  I., Imbriani,  G., \&  Piersanti,  L. 2003,
        \apj, 583, 878
\bibitem[Tassoul   et  al.(1990)]{1990ApJS...72..335T}   Tassoul,  M.,
        Fontaine, G., \& Winget, D.~E. 1990, \apjs, 72, 335
\bibitem[Torres   et   al.(2002)]{2002MNRAS.336..971T}   Torres,   S.,
        Garc{\'{\i}}a--Berro,  E., Burkert,  A., \&  Isern,  J.  2002,
        \mnras, 336, 971
\bibitem[Unglaub  \& Bues(2000)]{2000A&A...359.1042U} Unglaub,  K., \&
        Bues, I. 2000, \aap, 359, 1042
\bibitem[Vassiliadis  \& Wood(1993)]{1993ApJ...413..641V} Vassiliadis,
        E., \& Wood, P.~R.\ 1993, \apj, 413, 641
\bibitem[Weiss  \& Ferguson(2009)]{2009A&A...508.1343W} Weiss,  A., \&
        Ferguson, J.~W.\ 2009, \aap, 508, 1343
\bibitem[Winget  et   al.(1987)]{1987ApJ...315L..77W}  Winget,  D.~E.,
        Hansen,  C.~J., Liebert,  J., van  Horn, H.~M.,  Fontaine, G.,
        Nather,  R.~E., Kepler,  S.~O., \&  Lamb, D.~Q.\  1987, \apjl,
        315, L77
\bibitem[Winget  \& Kepler(2008)]{2008ARA&A..46..157W}  Winget, D.~E.,
        \& Kepler, S.~O. 2008, \araa, 46, 157
\bibitem[Wood(1992)]{1992ApJ...386..539W}  Wood,  M.~A.\  1992,  \apj,
        386, 539
\bibitem[Wood(1995)]{1995LNP...443...41W}  Wood, M.~A.\ 1995,  in {\sl
        ``White  Dwarfs''},  Eds: D.  Koester  \&  K. Werner,  Berlin:
        Springer-Verlag, Lecture Notes in Physics 443, 41
\end{thebibliography}
\end{document}